\documentclass[aps,pre,preprint,groupedaddress,showpacs]{revtex4-1}
\usepackage[dvips]{graphicx}
\usepackage{textcomp}
\usepackage{amssymb}

\newcommand{\mc}{\multicolumn}

\begin{document}

\title{
\Large\bf $\phi^4$ lattice model with cubic symmetry in three dimensions:
  RG-flow and first order phase transitions}

\author{Martin Hasenbusch}
\affiliation{
Institut f\"ur Theoretische Physik, Universit\"at Heidelberg,
Philosophenweg 19, 69120 Heidelberg, Germany}

\date{\today}
\begin{abstract}
We study the $3$-component $\phi^4$ model on the simple cubic lattice
in the presence of a cubic perturbation. To this end, we perform
Monte Carlo simulations in conjunction with a finite size scaling
analysis of the data.
The analysis of the renormalization group (RG)-flow of a 
dimensionless quantity provides us with the
accurate estimate $Y_4 - \omega_2 =0.00081(7)$ for the difference 
of the RG-eigenvalue $Y_4$ at the $O(3)$-symmetric fixed point and
the correction exponent $\omega_2$ at the cubic fixed point. 
We determine an effective exponent $\nu_{eff}$ of the correlation length
that depends on the strength of the breaking of the $O(3)$ symmetry.
Field theory predicts that depending on the sign of the
cubic perturbation, the 
RG-flow is attracted by the cubic fixed point, or runs to an ever 
increasing amplitude, indicating a fluctuation induced first 
order phase transition. We demonstrate directly the 
first order nature of the phase transition  for a sufficiently strong 
breaking of the $O(3)$ symmetry. We obtain accurate results for 
the latent heat, the correlation length in the disordered phase
at the transition temperature and the interface tension
for interfaces between one of the ordered phases and the 
disordered phase. We study how these quantities scale with the 
RG-flow, allowing quantitative predictions for weaker 
breaking of the $O(3)$ symmetry.
\end{abstract}

\maketitle
\section{Introduction}
We study the $\phi^4$ model with a cubic anisotropy in three dimensions. 
We focus on the case of $N=3$ components of the field, which is 
particularly interesting, since it is the experimentally most relevant case and
the cubic perturbation is very close to marginal at the $O(3)$-symmetric
fixed point. 
The model has been studied intensively over the last five decades, 
using field theoretic methods such as the $\epsilon$-expansion and perturbation 
theory in three dimensions fixed. For a review see for example Sec. 11.3 of
ref. \cite{PeVi02}.
Note that in structural transitions, in addition to $N=3$, 
$N=4$ might be experimentally realized \cite{Rong23}.
Recently the $\epsilon$-expansion has been extended to 6-loop \cite{epsilon6}. 
Based on this, a huge set of operator dimensions has been computed in
ref. \cite{BeHeKo23}.

In the field theoretic setting,
the reduced continuum Hamiltonian with two quartic couplings 
\begin{equation}
\label{ContHamil}
{\cal H} = \int \mbox{d}^d x  \left\{
\frac{1}{2} \sum_{i=1}^{N} [(\partial_{\mu}  \phi_i)^2
+r \phi_i^{2}]  + \frac{1}{4!} \sum_{i,j=1}^{N} (u + v \delta_{ij})
\phi_i^2  \phi_j^2   \right\}  \;,
\end{equation}
where $\phi_i$ is a real number, is studied,
(see for example eq.~(11.10) of ref. \cite{PeVi02}). Flow equations  in a
two-dimensional parameter space $(u,v)$ are discussed. For $v=0$, the theory 
is $O(N)$-symmetric, while for finite $v$, the theory has only cubic symmetry.
The qualitative features of the flow are well understood. 
There are four fixed points:
The Gaussian $(u,v)=(0,0)$, the decoupled Ising $(0,v^*)$, 
the $O(N)$-symmetric $(u,v)=(u^*,0)$  and the 
fixed point with cubic symmetry only $(u,v)=(u_c,v_c)$, where $v_c>0$.
The Gaussian and the decoupled Ising (DI) fixed points are unstable for all 
values 
of $N$ and $N>1$, respectively. The $O(N)$-symmetric fixed point is unstable 
for $N \ge N_c$ in one direction, breaking the $O(N)$ symmetry.
Recent field theoretic estimates give robustly $N_c$ slightly smaller than 3.
The result $N_c<3$ is supported by the fact that in a finite size
scaling analysis of Monte Carlo data for the improved $\phi^4$ model on the
simple cubic lattice the authors find $Y_4 = 0.013(4)$ for
$N=3$ \cite{O234}. In ref. \cite{myCubic} $Y_4=0.0142(6)$ had been obtained.
The rigorous bound $Y_4 > 3-2.99056$ for $N=3$ was recently established
by using the conformal bootstrap (CB) method \cite{CB_O3}. Note that $Y_4$ is 
the RG-exponent of the cubic perturbation at the $O(N)$-symmetric fixed point
and $Y_4>0$ means that the perturbation is
relevant and hence the RG fixed point is unstable. The cubic fixed point is 
stable for $N \ge N_C$ and for $v>0$ the flow runs into the cubic fixed point. 
On the contrary, for $v<0$, the flow runs to ever larger violations of the 
$O(N)$ symmetry and no fixed point is reached. Instead a fluctuation induced 
first order phase transition is expected.

In ref. \cite{AharonyNeu} the authors pointed out that the slow RG-flow, related
with the small value of $Y_4$, requires the knowledge of
the RG-flow also at finite distances from the fixed points in order to 
interpret experimental results. Slow in this context means that 
the couplings change only by little, when the length scale is varied over a 
range that can be studied in experiments or in simulations.
Just to get an 
idea: Take a sample of a material of linear size $10^{-1} \mbox{m}$ and a 
lattice constant of $10^{-10} \mbox{m}$. Then the amplitude of a cubic perturbation 
in the neighborhood of the $O(3)$-symmetric fixed point only increases by 
a factor of $(10^{9})^{0.0142}=1.34...$ going from the microscopic
to the macroscopic  scale, where we have taken the estimate of $Y_4$
for $N=3$ obtained in ref. \cite{myCubic}. In ref. \cite{AharonyNeu} it 
is demonstrated that the RG-flow rapidly collapses on a line connecting
the decoupled Ising, the cubic and the O(3)-symmetric fixed points.
Also for $v<0$, the RG-flow remains slow on the 
continuation of this line for a relatively large range of $v$. It follows 
that the
first order transition is very weak for a large range of the parameters, 
such that in experiments it appears as a continuous transition, albeit
with critical exponents slightly different from those of the 
$O(3)$-symmetric fixed point.
This picture has been worked out in ref. \cite{AharonyNeu} by using the 
coefficients of the 6-loop $\epsilon$-expansion obtained in 
ref. \cite{epsilon6}. For related work on systems with competing order
parameters by the same authors see \cite{AharonyNeu1,AharonyNeu2,AharonyNeu3}.
Here we corroborate and extend these results by using Monte Carlo simulations 
and finite size scaling (FSS) methods.

We build on ref. \cite{myCubic}, where 
we performed large scale Monte Carlo simulations of a 
lattice version of the $\phi^4$ model. We studied the cases $N=3$ and 
$4$, focussing on the neighborhood of the 
$O(N)$-symmetric and the cubic fixed points.  
By using finite size scaling, we computed accurate estimates of 
critical exponents for the cubic fixed point. In the case $N=3$ these differ 
only by little from their $O(N)$-symmetric counterparts.

In the present work, we extend the study of the RG-flow towards stronger 
violations
of the $O(3)$ symmetry. On the one hand, we make contact with the decoupled
Ising fixed point. On the other hand, for $v<0$, for large violations of the 
$O(3)$ symmetry, we demonstrate directly the first order nature of the 
transition. In our simulations we determine characteristic quantities such as 
the latent heat, the correlation length at the transition temperature,
and the interface tension between the disordered and one of the ordered phases.
We study how these quantities scale with the RG-flow. This way, quantitative
predictions can be made for all $v<0$. 

The outline of the paper is the following: In the next section we define
the model and the observables that we measure.
In section \ref{RGflow_basic} we discuss the basic ideas and objectives
of our study. The numerical results for the RG-flow are analyzed 
in section \ref{Numres}.
We determine effective critical exponents of the correlation length 
for different strengths of the symmetry breaking in section \ref{nueff}.
In section \ref{FirstOrder} we discuss our numerical results for the first 
order transition. In section \ref{summary} we summarize and conclude.
In appendix \ref{appendixA} we summarize technical aspects of our
numerical analysis and explain how we arrive at final results.
In appendix \ref{appendixB} we summarize basic analytic results derived 
from the $\beta$-function.

\section{The model and observables}
\label{Model}
Here we study the same reduced Hamiltonian and observables as in 
ref. \cite{myCubic}. For completeness let us recall the definitions.
We study a discretized version of the continuum Hamiltonian~(\ref{ContHamil}),
which is considered in field theory.
We extend the reduced Hamiltonian of the $\phi^4$ model on a simple cubic
lattice, see for example eq.~(1) of ref. \cite{ourHeisen},
by a term proportional to the traceless symmetric combination of four 
instances of the field, see for example eq.~(7) of ref. \cite{O234},
\begin{equation}
\label{spherical}
 \sum_{a} Q_{4,a a a a}(\vec{\phi}_x\,) = \sum_{a} \phi_{x,a}^4
-  \frac{3}{N+2}
\left( \vec{\phi}_x^{\,2} \right)^2 \;,
\end{equation}
with cubic symmetry, breaking $O(N)$ invariance.
Actually this
choice goes back to ref. \cite{Wegner}, where perturbations at the 
$O(N)$-symmetric fixed point were studied to leading order in the
$\epsilon$-expansion.
We get
\begin{equation}
\label{Hamiltonian}
 {\cal H}(\{\vec{\phi} \,\})= -\beta \sum_{<xy>} \vec{\phi}_x \cdot  \vec{\phi}_y
+ \sum_x \left [ \vec{\phi}_x^{\,2} + \lambda (\vec{\phi}_x^{\,2} -1)^2
+ \mu \left (\sum_{a} \phi_{x,a}^{\,4}
  -\frac{3}{N+2} \left( \vec{\phi}_x^{\,2} \right)^2 \right) \right]
\;,
\end{equation}
where 
$\vec{\phi}_x$ is a vector with $N$ real components.
The subscript $a$ denotes the components of the field and 
$\{\vec{\phi} \, \}$ is the collection of the fields at all sites $x$.
We label the sites of the simple cubic lattice by
$x=(x_0,x_1,x_2)$, where $x_i \in \{0,1,\ldots,L_i-1\}$. Furthermore,
$<xy>$ denotes a pair of nearest neighbors on the lattice.
In our study, the linear lattice size $L=L_0=L_1=L_2$ is equal in all 
three directions throughout. We employ periodic boundary conditions.
The real numbers $\beta$, $\lambda$ and $\mu$ are the parameters of the
model. Note that here $\lambda$ and $\mu$ take over the role of the 
parameters $u$ and $v$ of the continuum Hamiltonian, eq.~(\ref{ContHamil}). 
In the Hamiltonian~(\ref{Hamiltonian}) we take eq.~(\ref{spherical}) instead of 
simply $\sum_{a} \phi_{x,a}^4$, analogous to eq.~(\ref{ContHamil}), to achieve
\begin{equation}
\label{ortho} 
\left \langle A_{O(N)}(\{\vec{\phi} \, \}) \sum_{a,x} Q_{4,a a a a}(\vec{\phi}_x\,) 
\right \rangle_{\mu=0} = 0 \;,
\end{equation}
where the estimator $A_{O(N)}$ is $O(N)$ invariant, while for an estimator with 
cubic symmetry, breaking $O(N)$-invariance, we get in general a finite value.

\subsection{Decoupled systems}
In eq.~(\ref{Hamiltonian}) the components of the field 
decouple for $\lambda -\frac{3}{N+2} \mu=0$. Since the term 
$\sum_x \vec{\phi}_x^{\,2}$ has the factor $(1-2 \lambda)$ and 
$\sum_x \sum_a \phi_{x,a}^{4}$ the factor $\mu = \frac{N+2}{3} \lambda$
in front, a rescaling of the field $\phi_x$ is needed to match with the 
Hamiltonian 
\begin{equation}
\label{HamiltonianI}
 {\cal H}(\{\phi\})= -\tilde \beta \sum_{<xy>} \phi_x \phi_y
+ \sum_x \left [ \phi_x^{2} + \tilde \lambda (\phi_x^{2} -1)^2
   \right]
\;,
\end{equation}
considered for example in ref. \cite{myPhi4}, where $\phi_x$ is a real 
number. 
We arrive at the equations
\begin{equation}
 (1-2 \lambda)  = (1-2 \tilde \lambda) \; c \;\; ,\;\;\;\; 
\frac{N+2}{3} \lambda = \tilde \lambda \; c^2
\end{equation}
and hence
\begin{equation}
\frac{6}{N+2} \tilde \lambda \; c^2 +  (1-2 \tilde \lambda) \; c -1 =0 
\end{equation}
with the solutions
\begin{equation}
c =  \frac{-(1-2 \tilde \lambda) \pm \sqrt{(1-2 \tilde \lambda)^2 
 +\frac{24}{N+2} \tilde \lambda}}{\frac{12}{N+2} \tilde \lambda} \;\;,
\end{equation}
where we take the positive solution. Plugging in $\tilde \lambda^*=1.1(1)$
\cite{myPhi4} we arrive at $c=1.436(15)$ for $N=3$. Note that
$\tilde \lambda^*$ denotes the value of $\tilde \lambda$, where leading 
corrections to scaling vanish. Hence we get for the improved decoupled
model
$\lambda^*_{DI}=1.36(15)$ and $\mu^*_{DI} = \frac{N+2}{3} \lambda^*_{DI} =2.27(25)$.

\subsection{The observables and dimensionless quantities}
\label{observables}
Dimensionless quantities or phenomenological couplings play a central
role in finite size scaling. 
Similar to the study of $O(N)$-symmetric models, we study
the Binder cumulant $U_4$, the ratio of partition functions $Z_a/Z_p$ and
the second moment correlation length over the linear lattice size
$\xi_{2nd}/L$. Let us briefly recall the definitions of the observables
and dimensionless quantities that we measure.

The energy of a given field configuration is defined as
\begin{equation}
\label{energy}
 E=  \sum_{<xy>}  \vec{\phi}_x  \cdot \vec{\phi}_y \;\;.
\end{equation}

The magnetic susceptibility $\chi$ and the second moment correlation length
$\xi_{2nd}$ are defined as
\begin{equation}
\label{suscept}
\chi  =  \frac{1}{V}
\left\langle \Big(\sum_x \vec{\phi}_x \Big)^2 \right\rangle \;\;,
\end{equation}
where $V=L^3$ and
\begin{equation}
\xi_{2nd}  \equiv  \sqrt{\frac{\chi/F-1}{4 \sin^2 \pi/L}} \;\;,
\end{equation}
where
\begin{equation}
F  =  \frac{1}{V} \left \langle
\Big|\sum_x \exp\left(i \frac{2 \pi x_k}{L} \right)
        \vec{\phi}_x \Big|^2
\right \rangle
\end{equation}
is the Fourier transform of the correlation function at the lowest
non-zero momentum. In our simulations, we have measured $F$ for the three
directions $k=0,1,2$ and have averaged these three results.

The Binder cumulant $U_4$ is given by
\begin{equation}
U_{4} = \frac{\langle (\vec{m}^{2})^2 \rangle}{\langle \vec{m}^2\rangle^2},
\end{equation}
where $\vec{m} = \frac{1}{V} \, \sum_x \vec{\phi}_x$ is the
magnetization of a given field configuration. We also consider the ratio
$R_Z = Z_a/Z_p$ of
the partition function $Z_a$ of a system with anti-periodic boundary
conditions in one of the three directions and the partition function
$Z_p$ of a system with periodic boundary conditions in all directions.
This quantity is computed by using the cluster algorithm.
For a discussion see Appendix A 2 of ref. \cite{XY1}.

In order to detect the effect of the cubic anisotropy we
study
\begin{equation}
\label{UCdef}
U_C = \frac{\langle \sum_a Q_{4,aaaa}(\vec{m}) \rangle}
     {\langle \vec{m}^{\,2} \rangle^2}  \; .
\end{equation}
In the following we shall refer to the dimensionless
quantities $U_C$, $U_{4}$, $Z_a/Z_p$ and $\xi_{2nd}/L$ by
using the symbol $R$.
Note that $U_C=O(\mu)$, while $R=R|_{\mu=0} + O(\mu^2)$
for $U_4$, $Z_a/Z_p$, and $\xi_{2nd}/L$. 

In our analysis we need the observables as a function of $\beta$ in
some neighborhood of the simulation point $\beta_s$. To this end we have
computed the coefficients of the Taylor expansion of the observables
up to the third order.

\subsection{Dimensionless quantities for the  decoupled system and the 
first order transition}
\label{decouple_first}
In the case of decoupled one-component systems, 
$\lambda -\frac{N+2}{3} \mu=0$, we can express
the dimensionless quantities introduced above in terms of their one-component
counterparts.
For example 
\begin{equation}
\label{UCDI}
U_{C,DI} = \frac{N-1}{N (N+2)} (U_{4,Ising} - 3) \;\;,
\end{equation}
where $U_{4,Ising}$ is the Binder cumulant of the one-component
system. The calculation is straight forward, only exploiting that
$\langle m_a^2 m_b^2 \rangle = \langle m_a^2 \rangle  \langle  m_b^2 \rangle$
for $a \ne b$ for the decoupled case.
Hence we get for the fixed point value, which is indicated by $^*$
\begin{equation}
\label{UCDIS}
U_{C,DI}^*= (1.60359(4) -3)  \frac{N-1}{N (N+2)} = - 1.39641(4) 
\frac{N-1}{N (N+2)}
\end{equation}
using the result of \cite{myIso} for  $U_{4,Ising}^*$. 
Furthermore $(Z_a/Z_p)_{DI} = ((Z_a/Z_p)_{Ising})^N$, $U_{4,DI} = 
\frac{1}{N} U_{4,Ising} + \frac{N-1}{N}$, and 
$(\xi_{2nd}/L)_{DI} = (\xi_{2nd}/L)_{Ising}$, where the subscripts
$DI$ and $Ising$ indicate the decoupled and the one-component system, 
respectively.

For the first order transition, in the large $L$ limit, at the transition 
temperature, there is the disordered high temperature phase and the 
$2 N$-fold degenerate ordered phase. Each phase enters with the same 
weight \cite{BoKo90}. 
In the high temperature phase, the magnetization vanishes, while 
for the ordered phase there is a finite magnetization with the
modulus $m_{order}$. Fluctuations 
of the magnetization vanish as the lattice size is increased.
Hence $\langle \vec{m}^2 \rangle = \frac{0 + 2 N m_{order}^2}{1 + 2 N}=
\frac{2 N}{1 + 2 N} m_{order}^2$,
$\langle (\vec{m}^2)^2 \rangle = \frac{0 + 2 N m_{order}^4}{1 + 2 N}
=\frac{2 N}{1 + 2 N}  m_{order}^4$,
and $\langle \sum_a m_a^4 \rangle  = \frac{0 + 2 N m_{order}^4}{1 + 2 N}
=\frac{2 N}{1 + 2 N} m_{order}^4$.
Putting things together we get $U_4=\frac{2 N+1}{2 N}$ and 
$U_C=\frac{(2 N+1)(N-1)}{(2 N)(N+2)}$.
The ratio of partition functions assumes the value $Z_a/Z_p=0$ and $Z_a/Z_p=1$
in the limit $L \rightarrow \infty$ in the low and the high temperature phase,
respectively. Taking into account the degeneracy of the ordered  phase,
we arrive at $Z_a/Z_p=1/(2 N+1)$.

\section{Finite size scaling study of the RG-flow: theory}
\label{RGflow_basic}
Let us outline the general strategy of our analysis and 
discuss how it arises from the renormalization group theory.
Our theoretical framework is the realspace  RG
as it is discussed for example in ref. \cite{Barber}.
Furthermore we assume that the picture that arises from the field-theoretic
analysis \cite{AharonyNeu} is qualitatively correct.

The key assumption of finite size scaling is that the RG-flow is 
essentially unaffected by the finiteness of the system, until 
a scale close to the linear system size $L$ is reached. Furthermore,
the dimensionless quantities considered here are invariant under 
RG-transformations. Let us briefly discuss this point
for the cumulants $U_4$ and $U_C$. Both are constructed 
from the magnetization $\vec{m} = \frac{1}{V} \, \sum_x \vec{\phi}_x$.
The linear blockspin transformation is considered as a viable 
RG-transformation. It is given by
\begin{equation}
\label{linearT}
\vec{\Phi}_X = \frac{1}{b^{d-\Delta} } \sum_{x \in X} \vec{\phi}_x \;,
\end{equation}
where $b$, for example $b=2$, is the linear size of a $b^d$ block. 
The sites of the coarse lattice are denoted by $X$ and are
identified with blocks on the fine lattice. The field on the 
coarse lattice is denoted by $\vec{\Phi}_X$. The dimension of the 
field $\Delta$ is a priori unknown. Under the  
transformation~(\ref{linearT}), the magnetization of a configuration 
remains exactly the same up to the normalization $1/b^{d-\Delta}$.
The cumulants $U_4$ and $U_C$ are constructed such that this normalization 
cancels. Hence
$U_4$ and $U_C$ are exactly invariant under the linear blockspin 
transformation. For other types of RG-transformations this might be only 
approximately 
true. Therefore we expect that $U_4$ and $U_C$ are affected by corrections
related with the redundant RG-eigenvalues of the linear blockspin 
transformation. The leading one can be identified with the analytic background 
of the magnetic susceptibility. In contrast, the ratio of partition functions
$Z_a/Z_p$ should not be affected by such corrections, since partition functions
stay invariant under any type of RG-transformation. This is the reason why 
below we give preference to $Z_a/Z_p$, when $Z_a/Z_p$ or $\xi_{2nd}/L$ could 
be used. 

Typically, we study models with a few parameters. Here 
$\beta$, $\lambda$, and $\mu$.  The RG-flow starts with 
these parameters. Realspace RG-transformations generate an 
infinite number of couplings $K_{\alpha}$ already in the first 
step.
An important feature of the RG-flow is that it rapidly 
collapses onto low dimensional submanifolds. In the literature, 
mostly the collapse onto a fixed point is discussed. For our 
purpose, this has to be generalized. 
The RG-flow collapses onto lower and lower dimensional 
submanifolds in a hierarchical manner. In the neighborhood of 
fixed points, this hierarchy is given by the values of irrelevant
RG-eigenvalues $y_i$. In our problem, as seen in the analysis
of the 6-loop $\epsilon$-expansion \cite{AharonyNeu},
we have a rapid collapse on a line in coupling space. The 
RG-flow on this line is slow, corresponding to RG-eigenvalues 
$y \approx 0$ at the fixed points. Note that it is assumed that we are
on the critical surface.

In our numerical study, we do not compute the RG-flow in terms
of couplings $K_{\alpha}$. Instead, we monitor the  RG-flow by 
studying the behavior of dimensionless quantities $R_i$. 
Based on the invariance of dimensionless quantities under 
RG-transformations we might write them as a function of the transformed
couplings at the scale of the linear lattice size $L$:
\begin{equation}
 R(L,\beta,\lambda,\mu) = \hat R(\vec{K}(L,\beta,\lambda,\mu)) \;,
\end{equation}
where the hat indicates that $R$ and $\hat R$ are mathematically
different functions. The argument of $\hat R$ is defined as follows:
The RG-transformations start with our lattice model at
$(\beta,\lambda,\mu)$. Then $n$ blockspin transformations are performed,
such that $L=b^n$. The result of these $n$ transformations is
$\vec{K}(L,\beta,\lambda,\mu)$. For sufficiently large $L$, the RG-flow
collapses on a line, up to small corrections. Therefore we might 
write
\begin{equation}
\hat R(\vec{K}(L,\beta=\beta_c,\lambda,\mu))=
\tilde R(\tilde v(L,\beta=\beta_c,\lambda,\mu)) \;.
\end{equation}
We explicitly indicate that we are on the critical surface by
setting $\beta=\beta_c$ and the real $\tilde v$ parametrizes the 
line of the slow flow. $\tilde v$ should be an analytic function of 
$\vec{K}$ and $\tilde v(L,\beta=\beta_c,\lambda,\mu)=O(\mu)$.
These requirements are actually fulfilled by the  dimensionless 
quantity $U_C$, which we consider as coupling in the following. Roughly
speaking, the physics at large scales depends only on the strength of
the breaking of the $O(N)$ symmetry, and this strength can be monitored
by using $U_C$. Note, to avoid confusion later on, that $U_C$ has the 
opposite sign as  $\mu$.
Our approach is inspired by ref. \cite{running},
where the dimensionless quantity $\xi(L)/L$ is considered
as coupling in an asymptotically free theory. In ref. \cite{running} a 
lattice of the size $L \times \infty$ is considered and $\xi(L)$ is the 
exponential correlation length. For a complementary point of view see ref.
\cite{Cara95}.
Similar to ref. \cite{running}, an RG-transformation by the scale factor $b$
is given by
\begin{equation}
\label{RGtrafo}
U_C'(L,\beta=\beta_c,\lambda,\mu) = U_C(b L,\beta=\beta_c,\lambda,\mu)
\end{equation}
for large $L$. The fixed points are:
For $O(N)$ symmetry, at $\mu=0$, $U_C'=U_C=0$, at the cubic fixed point
$U_C'=U_C=U_C^*$, which we compute below, and at the decoupled Ising 
fixed point $U_C'=U_C=U_{C,DI}^*$, eq.~(\ref{UCDIS}).
Furthermore for dimensionless quantities $R$ different from $U_C$,
\begin{equation}
\label{otherR}
R(L,\beta=\beta_c,\lambda,\mu)= \tilde R(U_C(L,\beta=\beta_c,\lambda,\mu))
\end{equation}
for large $L$. The large $L$ or scaling limit is obtained 
by approaching the unstable fixed points, keeping $U_C$ fixed. At the 
$O(N)$-invariant fixed point this means that we have to take the limit
$L \rightarrow \infty$ and $\mu \rightarrow 0$ simultaneously, such that
$U_C$ stays fixed. The analysis 
of the 6-loop $\epsilon$-expansion \cite{AharonyNeu} suggests that the 
collapse of the RG-flow on a single line still occurs far off from the 
fixed points. Therefore in our numerical study we expect to see a good
approximation of the scaling limit already for moderately large $L$, while 
$\mu$ is still quite different from zero.

In our numerical study, we compute the RG-transformation~(\ref{RGtrafo})
for finite $b$. For convenience, we view this as a finite difference
approximation of a $\beta$-function, where infinitesimal changes of 
$\ln L$ are considered:
\begin{equation}
\label{infinitesimal_flow}
\tilde u(U_C) = \frac{\mbox{d} U_C}{\mbox{d} \ln L} \approx 
\frac{U_C(b L) - U_C(L)}{\ln (b L)- \ln L}
= \frac{U_C(b L) - U_C(L)}{\ln b}  \; \;.
\end{equation}
At least for $U_C$ not too large, the RG-flow is slow and well behaved. 
Therefore this should be a good approximation.

In our study we investigate the flow in the whole range of interest.
On the one hand we start at the decoupled Ising fixed point. 
The value $U_{C,DI}^*$ can be expressed in terms of $U_4^*$ of the 
Ising universality class, which is known to high precision \cite{myIso}.
The relevant RG-exponent at this fixed point can be expressed in terms 
of the thermal RG-exponent of the Ising universality class 
\cite{Sak74,Carmona}. The flow runs towards the cubic fixed point.
This part of the study can be seen as a benchmark of our approach.
Then there is the crossover from the $O(N)$-symmetric fixed point 
to the cubic fixed point. Our main focus is on the flow away from the 
$O(N)$-symmetric fixed point, towards increasing values of $U_C$, 
with a first order phase transition of increasing strength.

Simulating at a given pair $(\lambda,\mu)$ we can follow the 
RG-flow by simply increasing the lattice size $L$.  However, since the 
RG-flow is slow, only a small range in $U_C$ can be accessed this way.  
In order to study the whole range, simulations for many different pairs
$(\lambda,\mu)$ have to be performed. The results of these simulations are 
patched together. For moderate sizes of $|U_C|$ this is done by fitting
the numerical estimates of $u= \frac{1}{U_C} \tilde u$ by using a polynomial
Ansatz. 
Note that the zero of $\tilde u$ at $U_C=0$ is lifted in $u$.
The larger $U_C>0$, the worse is
approximation~(\ref{infinitesimal_flow}) and also fitting $u$
by using a polynomial Ansatz requires more and more orders. Therefore,
for larger $U_C$, the patching is done more directly by matching the linear
lattice sizes for two pairs of $(\lambda,\mu)$ such that the estimates of 
$U_C$ fall on top of each other:
\begin{equation}
\label{matching}
U_C(L,\beta=\beta_c,\lambda_1,\mu_1) = 
U_C(c_{1,2} L,\beta=\beta_c,\lambda_2,\mu_2)  \;.
\end{equation}
For $\mu_1, \mu_2<0$, we can not expect that there is a 
$L \rightarrow \infty$ limit of $c_{1,2}$, since eventually we see the first 
order phase transition. However, we get estimates of $c_{1,2}$ that are
consistent at the level of our numerical precision for some range in $L$.

Let us summarize the objectives of the study: 
\begin{itemize}
\item Numerically check that the RG-flow collapses on a line by, for example,
verifying eq.~(\ref{otherR}). This done below in Sec. \ref{UCflow}.

\item Numerically determine the $\beta$-function, 
eq.~(\ref{infinitesimal_flow}). Based on this we compute the fixed point
value $U_C^*$ at the cubic fixed point, the RG-exponent $Y_4$ at the 
$O(3)$-symmetric fixed point and the correction exponent $\omega_2$ at
the cubic fixed point. Note that the exponents are obtained from the 
derivative of the $\beta$-function at the fixed points with respect to
$U_C$. This is done below in Sec. \ref{UCflow}.

\item Motivated by ref. \cite{AharonyNeu} we compute in Sec. \ref{nueff}
effective critical exponents as a function of $U_C$. This is done by 
using finite size scaling, and as a check by using the scaling of
the correlation length in the thermodynamic limit.

\item Finally, in Sec. \ref{FirstOrder} we demonstrate explicitly the first 
order nature of the transition for $\mu < 0$ and $|\mu|$ sufficiently large.
In particular, we compute the correlation length $\xi_{high}$ in the high 
temperature phase at the transition temperature in the thermodynamic 
limit with high accuracy. Based on the scaling Ansatz
\begin{equation}
\xi_{high}(\beta=\beta_c,\lambda,\mu) =  
\hat \xi_{high}(U_C(L,\beta=\beta_c,\lambda,\mu)) \; L
\end{equation}
we obtain $\xi_{high}$ for $(\lambda,\mu)$, where we can not simulate
lattices with $L \gg \xi_{high}$, which is required to get a good 
approximation of the thermodynamic limit, using our limited computational
resources.  
Here we denote, a bit sloppy, 
the inverse transition temperature by $\beta_c$. In the spirit of 
ref. \cite{Cara95}, by using
$\xi_{high}(\lambda_2,\mu_2)=c_{1,2} \xi_{high}(\lambda_1,\mu_1)$, we obtain
iteratively results for weaker and weaker transitions, where the 
factor $c_{1,2}$ is given by eq.~(\ref{matching}). The latter is already
done in Sec. \ref{UCflow} using results obtained in Sec. \ref{FirstOrder}.
\end{itemize}

\section{Finite size scaling study of the RG-flow: Numerics}
\label{Numres}
As first step of our numerical study, 
we repeat the analysis of section VII of ref. \cite{myCubic}
with more data. We have added new pairs of $(\lambda,\mu)$, in particular
for relatively large values of $|\mu|$. Furthermore,
for pairs $(\lambda,\mu)$ already studied in ref. \cite{myCubic}
we improved the statistics and added larger lattice sizes. As a preliminary
step, generalizing the idea of improved models, 
we locate the line of slow flow in the $(\lambda,\mu)$ plane. 
Then we continue with the study of the flow of $U_C$ as outlined in 
the previous section. Note that in appendix \ref{appendixA} we 
collect some technical aspects of the numerical analysis and discuss
how we arrive at final results.

\subsection{Locating the line of slow RG-flow in the $(\lambda,\mu)$ plane}
\label{slowline}
In the study of critical phenomena by using lattice models, improved models 
have been demonstrated to be helpful in obtaining accurate estimates of 
universal quantities. For a discussion, 
see for example Sec. 2.3 of ref. \cite{PeVi02}. In the case of 
the Heisenberg universality class, improved models had been studied for 
example in refs. \cite{ourHeisen,O234,myIco}. The basic idea is that a
parameter of the model is tuned such that the scaling field of the leading 
correction to scaling vanishes. In the case of the 3-component 
$\phi^4$ model, eq.~(\ref{Hamiltonian}) with $\mu=0$, this is achieved 
for $\lambda^*=5.17(11)$, see ref. \cite{myIco}. 
In ref. \cite{myCubic}, we have 
applied the idea to the cubic fixed point, now eliminating the scaling fields
of the two leading corrections. For $N=3$, we get 
$(\lambda,\mu)^*_{cubic}=(4.99(11), 0.28(2))$. Above in section
\ref{Model}, we have already discussed how the result \cite{myPhi4} 
for the one component $\phi^4$ model translates to the decoupled $\phi^4$ case.
In the case of the $O(3)$-symmetric $\phi^4$ model, we have analyzed our 
data for dimensionless quantities by using Ans\"atze of the form
\begin{equation}
\label{Ransatz0}
R_i(\beta_c,\lambda,L) = R_i^* + r_{i}  w(\lambda) L^{-\omega} + ... \;,
\end{equation}
where $\lambda^*$ is obtained as zero of $w(\lambda)$. In ref. \cite{myCubic},
eq.~(66), we have generalized eq.~(\ref{Ransatz0}) to the line of slow flow
that we eventually intend to study:
\begin{equation}
\label{R3master}
R_i(\beta_c,\lambda,\mu,L) =R_i^*
+ \sum_{m=2}^{m_{max}} c_{i,m} U_C^m(\beta_c,\lambda,\mu,L)
+  r_{i}  w(\lambda,\mu) L^{-\omega}
+ \sum_j a_{i,j} L^{-\epsilon_j} 
\end{equation}
for dimensionless quantities that behave like
$R_i(\beta_c,\lambda,\mu,L)=R_i(\beta_c,\lambda,0,L) + O(\mu^2)$. 
The $O(\mu^2)$ contributions are taken into account by the term
$\sum_{m=2}^{m_{max}} c_{i,m} U_C^m(\beta_c,\lambda,\mu,L)$. 
In fact, we perform a Taylor expansion of eq.~(\ref{otherR}).
Note that we obtain dimensionless 
quantities $R_i-\sum_{m=2}^{m_{max}} c_{i,m} U_C^m$ that, at least
approximately, remain invariant under the slow part of the RG-flow. 
Corrections are assumed to be of the same form as corrections to scaling
in the neighborhood of a fixed point. Furthermore,
the exponent $\omega$ is assumed to be constant, taking the value 
$\omega=0.759(2)$ of the Heisenberg universality class \cite{myIco}.
This is motivated by the fact that
$\omega_1$ of the cubic universality class differs by little from $\omega$ 
of the Heisenberg one. Furthermore $\omega$ of the Ising universality class,
which is the correction exponent at the decoupled Ising fixed point,
is only slightly larger than the correction exponent of the Heisenberg 
universality class. In our fits we set $r_{U_4}=1$, while $r_{Z_a/Z_p}$ and
$r_{\xi_{2nd}/L}$ are free parameters. Terms  proportional to $L^{-2 \omega}$
and higher are not taken into account, since for our data the correction
is small. The last term of eq.~(\ref{R3master}) takes into account subleading
corrections with $\epsilon_j \gtrapprox 2$.  Also these correction 
exponents are assumed to be constant. In the fits performed here,
we took two exponents $\epsilon_1=2-\eta$ and $\epsilon_2=2.023$, corresponding
to the analytic background of the magnetization and the breaking of the 
rotational invariance by the lattice. In the case of the analytic background, 
we assume that the coefficients depend linearly on $\lambda$ and quadratically
on $\mu$. The coefficients for the breaking of the rotational invariance are
taken as a constant.  Formally, in eq.~(\ref{R3master}), the correction term
$r_{i}  w(\lambda,\mu) L^{-\omega}$ is of the same type as those contained in
$\sum_j a_{i,j} L^{-\epsilon_j}$.  We single out the term with the 
smallest correction exponent, since in the following we like to focus on
$(\lambda,\mu)$, where $w(\lambda,\mu) \approx 0$, which defines the 
line of the slow flow in the $(\lambda,\mu)$ plane. 

The Ansatz~(\ref{R3master}) is used for joint fits of $Z_a/Z_p$, 
$\xi_{2nd}/L$, and $U_4$.
In a first series of  fits, we took $w(\lambda,\mu)$ as a free parameter
for each value of $(\lambda,\mu)$. We performed a number of fits, varying 
the range of $\mu$ that is taken into account, the maximal power 
$m_{max}$ of $U_C$ and, as usual, the minimal linear lattice size $L_{min}$
taken into account. 
The different sets of data that we analyzed are mainly characterized by the 
range of $\mu$ that is taken. For the smallest set of data, $|\mu| \le 1.2$ 
is taken, while for the largest $-1.8 \le \mu \le 2.2$ is taken.  The largest
set contains 64 different pairs of $(\lambda,\mu)$. 
For negative $\mu$ we used the additional cut $U_C \lessapprox 0.4$. 
As a result, for $(\lambda,\mu)=(3.4,-1.8)$, $(3.0,-1.663)$ and $(2.7,-1.552)$
only linear lattice sizes up to $L=48$ are used in the fit.

In the case of our largest data set, we performed fits with $m_{max} \le 9$.  
For $m_{max}=9$, we get $\chi^2/$DOF$=1.070$, $1.045$, and $1.002$,
corresponding to $p=0.049$, $0.154$, and $0.478$
for $L_{min}=20$, $24$, and $28$, respectively.  For the definitions of 
$\chi^2/$DOF and the $p$-value see appendix \ref{appendixA}.  
For smaller ranges of $\mu$, acceptable fits were obtained already for smaller 
$L_{min}$. For example for $L_{min}=16$,  $|\mu| \le 1.2$, and $m_{max}=6$ 
we get $\chi^2/$DOF$=1.052$, corresponding to $p=0.103$.
Note that in ref. \cite{myCubic} we have used $m_{max}=5$ at most and 
fitted data for $|\mu| \le 1$.

Next we used the parameterizations 
\begin{equation}
\label{favoritew4}
 w(\lambda,\mu) = a (\lambda - \lambda^* - c \mu^2 -d  \mu^3) \;
  (1+e (\lambda-5.0))
\end{equation}
and
\begin{equation}
\label{favoritew4x}
 w(\lambda,\mu) = a (\lambda - \lambda^* - c \mu^2 -d  \mu^3 -e \mu^4) \;
  (1+f (\lambda-5.0) ) 
\end{equation}
for the correction amplitude.

Here acceptable fits were only obtained for data sets with a range 
up to $1.5 \ge \mu \ge -1.566$.
Various acceptable fits, using eq.~(\ref{favoritew4x}), are consistent with 
the estimates
$\lambda^* = 5.12(5)$, $c=-0.8(1)$, $d=0.06(1)$, and $e=0.01(4)$ for the 
line of slow flow.

\begin{figure}
\begin{center}
\includegraphics[width=14.5cm]{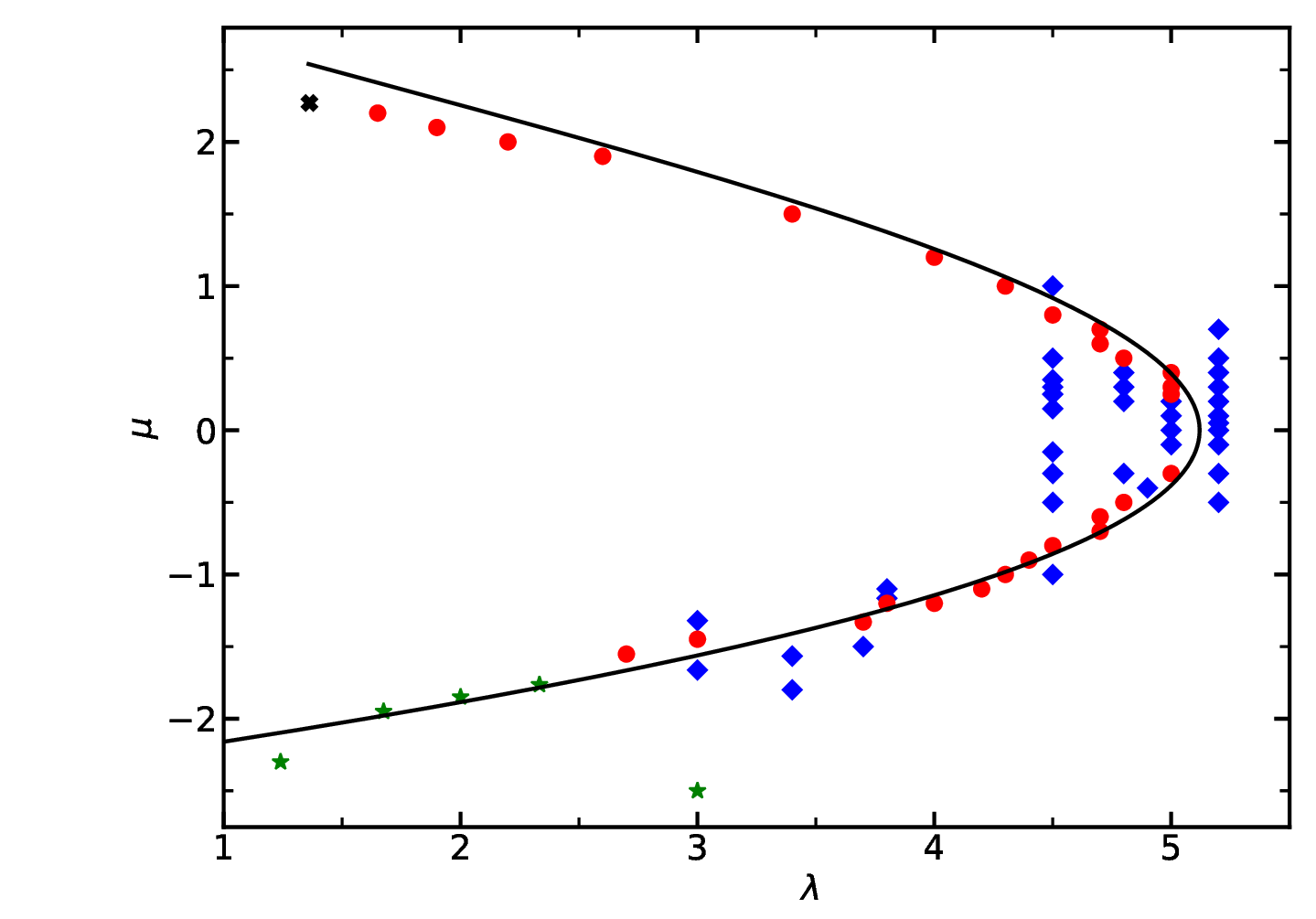}
\caption{\label{lambdamu}
We plot the line of slow flow in the $(\lambda,\mu)$ plane 
as parameterized by eq.~(\ref{favoritew4x}) using
the numerical values $\lambda^* = 5.12$, $c=-0.8$, $d=0.06$, and $e=0.01$
as solid black line. 
The solid circles give values of $(\lambda,\mu)$ with a small correction 
amplitude $w$. These are used in the analysis of the RG-flow below.
Additional  pairs of $(\lambda,\mu)$ that are analyzed in this section 
are plotted as diamonds.
The cross gives the improved point for the decoupled Ising system. 
The asterisks give points, where, below in Sec. \ref{FirstOrder},  we demonstrate
directly that the transition is first order.
}
\end{center}
\end{figure}

In Fig. \ref{lambdamu}  we plot the line of slow flow as characterized
by eq.~(\ref{favoritew4x}) with the numerical values of the parameters
given above. In addition we plot the pairs of $(\lambda,\mu)$ we have simulated
at. The pairs of $(\lambda,\mu)$ with a small
correction amplitude $w$ are shown as solid circles.
Here, small means that in fits without 
parameterization of $w$, the modulus of the value of $w$ is at most a few 
times the error of $w$. The improved point of the decoupled Ising system
is obtained from $\tilde \lambda^*=1.1(1)$ for the one-component
$\phi^4$ model on the simple cubic lattice \cite{myPhi4}
as discussed in section \ref{Model}.
Finally the pairs of $(\lambda,\mu)$, where, below in section \ref{FirstOrder}, 
we demonstrate directly that the transition is first order, are plotted.

Our results for the dimensionless quantities are fully consistent with those 
obtained in ref. \cite{myCubic}.  The estimates of $c_{i,m}$, summarized 
in table \ref{coeff_c}, are more accurate now. The error bars are taken 
such that the results of five different acceptable fits are covered.
We give results up to $m=4$. For larger $m$, 
the values of $c_{i,m}$ differ substantially between different fits.

\begin{table}
\caption{\sl \label{coeff_c}
Estimates of the coefficients $c_{i,m}$, eq.~(\ref{R3master}), for the three
dimensionless quantities that we have analyzed. For a discussion see the text.
}
\begin{center}
\begin{tabular}{cccc}
\hline
 $R_i$ $\backslash$ $m$ & 2 & 3 & 4 \\
\hline
$Z_a/Z_p$     & -0.612(9) &  2.13(8) & -5.0(1.2) \\  
$\xi_{2nd}/L$ & 1.308(9)  & -3.38(7) & 11.7(9) \\
$U_4$       & 1.243(8)  & -2.73(6) & 8.6(1.4) \\
\hline
\end{tabular}
\end{center}
\end{table}

Selected estimates of the inverse transition temperature 
$\beta_c$ are provided as supplementary material \cite{betacvalues}. 

\subsection{Flow equation for $U_C$}
\label{UCflow}
Now we proceed with the analysis of the RG-flow, as outlined 
in section \ref{RGflow_basic}. We like to study the RG-flow on the 
critical surface. To this end, following ref. \cite{HaPiVi}, we take the 
quantities of interest at $\beta=\beta_f$, where a dimensionless 
quantity $R_i$, different from $U_C$, assumes a certain value 
$R_{i,f}$: $R_i(\beta_f)=R_{i,f}$. Quantities taken at $\beta_f$ are
indicated by a line on top. For example $U_C$ at $\beta_f$ is
denoted by $\overline{U}_C$ in the following. Replacing a 
numerical estimate of $\beta_c$ by $\beta_f$ has various
advantages. For example, the statistical error is typically reduced due 
to cross-correlations with $R_i$, see for example ref. \cite{Parisen}.
Following previous work, for example \cite{myPhi4,XY1,ourHeisen,myIco}, 
we take the ratio of partition functions $Z_a/Z_p$ to define $\beta_f$. Our
choice for the fixed value is $(Z_a/Z_p)_f=0.19477$, which is the 
estimate of the fixed point value for the Heisenberg universality class
\cite{myIco}. Note that for any value of $R_{i,f}$ in the range of $R_{i}$,
for a second order phase transition, $\beta_f$ converges to $\beta_c$ 
as the linear lattice size $L$ increases. For $R_{i,f}=R_{i}^*$ the convergence
is the fastest.

Motivated by the results of the previous subsection, we consider as alternative
\begin{equation}
\label{tildeR}
 (Z_a/Z_p)(\beta_f) - \sum_{j=2}^{m_{max}} c_{Z_a/Z_p,j} U_C^j(\beta_f) 
= 0.19477 \;,
\end{equation}
where $c_{Z_a/Z_p,j}$ and $m_{max}$ are fixed. The idea of using 
eq.~(\ref{tildeR}) 
is that, in particular for large values of $|U_C|$, the convergence 
of $\beta_f$ with increasing $L$ is improved by using a quantity 
that is approximately invariant under the RG-flow.
In our numerical analysis we use $m_{max}=6$.
The values $c_{Z_a/Z_p,2}=-0.61$ and $c_{Z_a/Z_p,3}=2.1$ are taken 
from the fits discussed above, while
$c_{Z_a/Z_p,4}=-2.9$, $c_{Z_a/Z_p,5}=-10$, and $c_{Z_a/Z_p,6}=20.8$
are chosen such that for large $|U_C|$ certain requirements to 
be discussed below are fulfilled. We do not simply use numerical results
of $c_{Z_a/Z_p,j}$ for $j>3$, obtained by the fits performed in the previous
section, since for $U_C$ outside of the range of these fits the invariance
under the RG-flow is rapidly lost. Instead, the coefficients are chosen such
that the decoupled Ising fixed point and the first order transition
on the other side are approached properly. 

To this end, let us view
\begin{equation}
(Z_a/Z_p)_{mod,f}(U_C) = 
\sum_{j=2}^{m_{max}} c_{Z_a/Z_p,j} U_C^j +  (Z_a/Z_p)_{f} \;
\end{equation}
as a function of $U_C$. The coefficients 
$c_{Z_a/Z_p,4}$, $c_{Z_a/Z_p,5}$, and $c_{Z_a/Z_p,6}$  in eq.~(\ref{tildeR})
are chosen such that
$(Z_a/Z_p)_{mod,f}(U_C)$
\begin{itemize}
\item
is monotonically 
decreasing with increasing $U_C$ in the range $0 < U_C \lessapprox 0.4$. 
\item
Assumes roughly the numerical value found for $Z_a/Z_p$ at the transition 
temperature for $U_C \approx 0.4$. 
\item
is monotonically decreasing with decreasing $U_C$ in the range $0 > U_C \ge
U_{C,DI}$.
\item
Assumes the decoupled Ising value of $Z_a/Z_p$ for $U_{C,DI}$.
\end{itemize}

One should note that the numerical results discussed below in 
Sec. \ref{FirstOrder} show that at the inverse
transition temperature $\beta_t$, with increasing linear lattice size $L$,
the limiting values of the dimensionless quantities are not approached
monotonically. For example, one finds the extrema 
$(Z_a/Z_p)_{min}\approx 0.131$ and $U_{C,max} \approx 0.504$, both at 
$L/\xi_{high} \approx 2$, consistently for different values
of $\xi_{high}$. Plugging in $N=3$ into the equations of section 
\ref{decouple_first}, we get $Z_a/Z_p=1/7=0.142857...$ and 
$U_{C}=0.466666...$ in the large $L$ limit.
Studying the flow of $\overline{U}_C$, we stay in the range of $L$, 
where it is monotonic.

We performed the following analysis for both choices $(Z_a/Z_p)_f=0.19477$ and
eq.~(\ref{tildeR}). We obtain consistent results for these two choices. 
It is assuring that the results virtually do not depend on the choices, partly
made ad hoc, when fixing $c_{Z_a/Z_p,j}$ in eq.~(\ref{tildeR}). Not to 
overburden the reader, 
in the discussion below we only present the results obtained 
by using eq.~(\ref{tildeR}).
In order to keep corrections to scaling small, we focus on data obtained 
for $(\lambda,\mu)$ close to the line of slow flow 
as discussed above in section \ref{slowline}.

\begin{figure}
\begin{center}
\includegraphics[width=14.5cm]{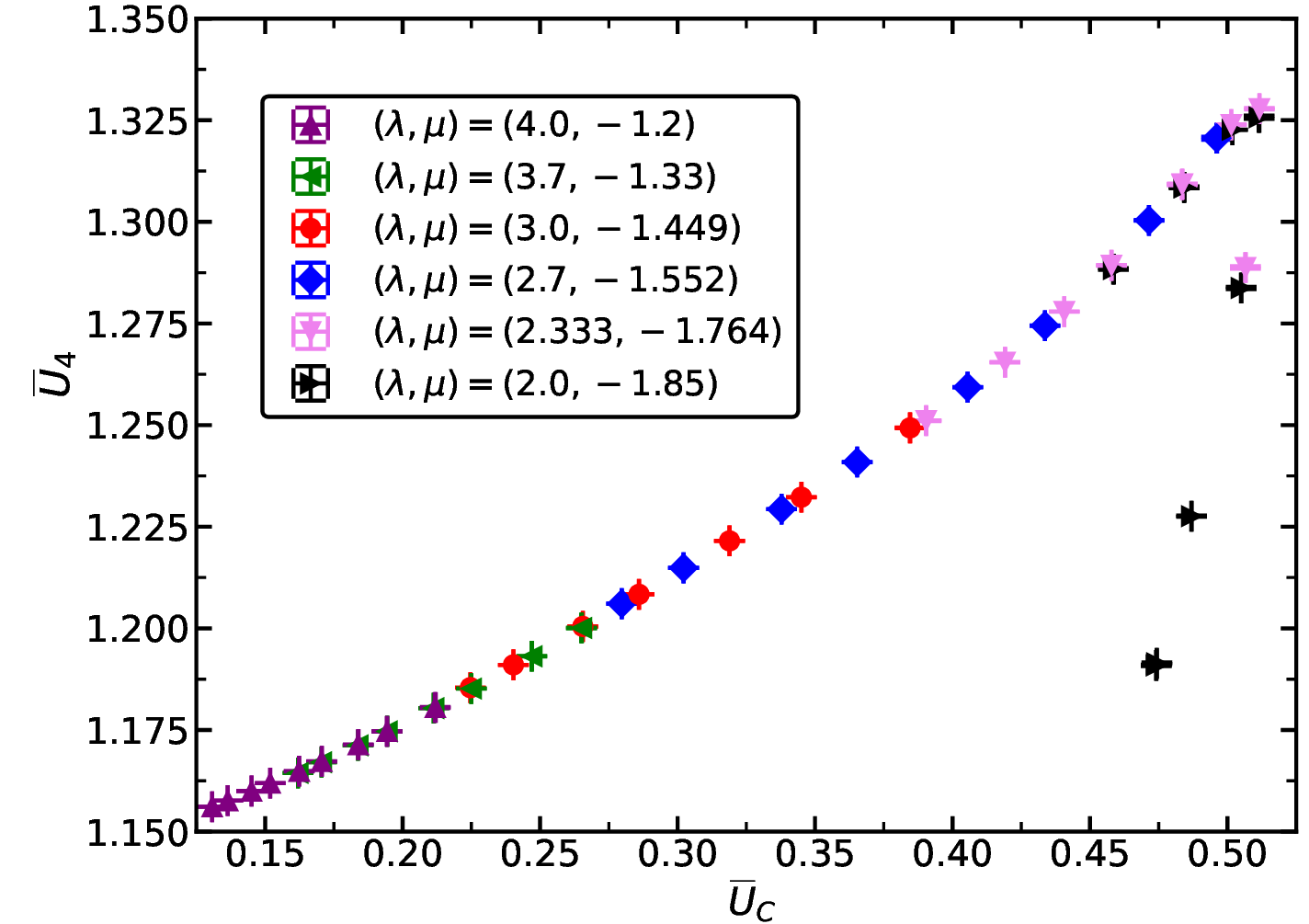}
\caption{\label{binderplot}
We plot $\overline{U}_4$ versus $\overline{U}_C$ obtained for six different 
pairs of $(\lambda,\mu)$ and a number of linear lattice sizes $L \ge 12$.
The largest lattice size is $L=192$, $128$, $96$, $128$, $64$, and $48$   
for 
$(\lambda,\mu)$=$(4.0,-1.2)$, $(3.7,-1.33)$, $(3.0,-1.449)$, $(2.7,-1.552)$,
$(2.333,-1.764)$, and $(2.0,-1.85)$, respectively. For a discussion see
the text.
}
\end{center}
\end{figure}

In order to check eq.~(\ref{otherR}), we plot $\overline{U}_4$ versus 
$\overline{U}_C$ for six pairs of $(\lambda,\mu)$ in Fig.~(\ref{binderplot}).
For $(2.333,-1.764)$ and $(2.0,-1.85)$, we show below in section 
\ref{FirstOrder} that the transition is of first order. The data points
for the different  pairs of $(\lambda,\mu)$ fall nicely on a single curve, 
confirming that there is a single parameter RG-flow, which is reached 
to a good approximation for the linear lattice sizes $L \ge 12$ studied here.
For $(2.333,-1.764)$ and $(2.0,-1.85)$ we see a second branch of the curve,
which is due to the non-monotonic behavior discussed above. Even here the
data for $(\lambda,\mu)=(2.333,-1.764)$ and $(2.0,-1.85)$ seem to fall 
on top of each other.

The flow of $\overline{U}_C$ is characterized by
\begin{equation}
\label{flowequation}
u(\overline{U}_C)= 
\frac{1}{\overline{U}_C}  \frac{\mbox{d} \overline{U}_C}{\mbox{d} \ln L} 
  \;,
\end{equation}
where we have introduced the factor $\frac{1}{\overline{U}_C}$, compared with 
a $\beta$-function in field theory, for numerical convenience. 
The factor lifts the zero of the $\beta$-function at $\overline{U}_C=0$.
Note that
here we replace $U_C(L,\beta=\beta_c,\lambda,\mu)$ of section 
\ref{RGflow_basic} by $U_C(L,\beta=\beta_f,\lambda,\mu)$.
Let us first discuss the behavior of $u$ 
in the neighborhood of the decoupled Ising fixed point. 
The decoupled Ising fixed point is unstable with an RG-exponent 
\begin{equation}
y = \alpha_I y_{t,I} = 2 y_{t,I} - d = 
d - 2 \Delta_{\epsilon, I} = 0.17475(2) \;,
\end{equation}
where $\alpha_I$ and $y_{t,I}$ are the specific heat and the thermal 
RG-exponent of the three-dimensional 
Ising universality class, respectively \cite{Sak74,Carmona}.
The numerical value of the scaling dimension 
$\Delta_{\epsilon} = 1.412625(10)$  is taken from ref.
\cite{Kos:2016ysd}. In the neighborhood of 
the decoupled Ising fixed point we have
\begin{equation}
\overline{U}_{C}(L) = \overline{U}_{C,DI} + \epsilon_0 (L/L_0)^y + ... \;.
\end{equation}
Hence 
\begin{equation}
\label{u_nextDI}
u(\overline{U}_{C,DI} +\epsilon_0) \approx 
\overline{U}_{C,DI}^{-1}  \lim_{L \rightarrow L_0}
\frac{\epsilon_0 (L/L_0)^y-\epsilon_0}{\Delta \ln L} = \overline{U}_{C,DI}^{-1}
\lim_{L \rightarrow L_0}  \frac{\epsilon_0 y \Delta \ln L}{\Delta  \ln L}
= \overline{U}_{C,DI}^{-1} \epsilon_0 y \;,
\end{equation}
where  $\Delta \ln L =\ln L - \ln L_0 =  \ln(L/L_0)$. Furthermore, we have used 
$(L/L_0)^y=\exp(y \ln (L/L_0))= 1 + y \ln (L/L_0) + ...$ .

In ref. \cite{myCubic} we estimated
$u$, eq.~(\ref{flowequation}), by fitting data for fixed $(\lambda,\mu)$ 
by using the Ansatz
\begin{equation}
\label{mostsimple}
\overline{U}_C(\lambda,\mu,L)  = a L^{u}
\end{equation}
or as check
\begin{equation}
\label{mostsimple2}
\overline{U}_C(\lambda,\mu,L)  = a L^{u} \;(1+c L^{-2})
\end{equation}
for some range $L_{min} \le L \le  L_{max}$. As argument of $u$ we took
$[\overline{U}_C(L_{min}) + \overline{U}_C(L_{max})]/2$. 
The approximations~(\ref{mostsimple},\ref{mostsimple2})
rely on the fact that $\overline{U}_C$ varies only little
in the range $L_{min} \le L \le L_{max}$ and hence $u$ is small:
\begin{eqnarray}
\overline{U}_C(L)  & \approx & \overline{U}_C(L_{0}) + 
\overline{U}_C(L_{0}) \; u \ln(L/L_{0}) 
= 
\overline{U}_C(L_{0}) \; [1+ u \ln(L/L_{0})] \nonumber \\
&\approx &
\overline{U}_C(L_{0}) \; \exp[u \ln(L/L_{0})] \propto L^{u} \;,
\end{eqnarray}
where $L_0 = \sqrt{L_{min} L_{max}}$.  In eq.~(\ref{mostsimple2}), 
the factor $(1+c L^{-2})$ is included to take subleading corrections
to scaling approximately into account.

For $(\lambda,\mu)$, where $\overline{U}_C$ changes considerably 
over the range of lattice sizes $L$ that we simulate, we now take instead
\begin{equation}
\label{diffdef}
 u([\overline{U}_C(L_2)+\overline{U}_C(L_1)]/2) = 
\frac{2}{\overline{U}_C(L_2)+\overline{U}_C(L_1)} 
\frac{\overline{U}_C(L_2)-\overline{U}_C(L_1)}{\ln(L_2/L_1)} 
\end{equation}
as approximation. Here $L_1$ and $L_2$ are lattice sizes we simulated at and 
$L_2$ is the smallest that satisfies $L_2 > L_1$, meaning the next larger to
$L_1$. Note that eq.~(\ref{diffdef})
is the finite difference approximation of eq.~(\ref{flowequation}).

First we check that estimates of $u$, 
eq.~(\ref{flowequation}),
obtained for different values of $(\lambda,\mu)$ fall on a unique curve,
up to small deviations that can be interpreted as corrections. 

\begin{figure}
\begin{center}
\includegraphics[width=14.5cm]{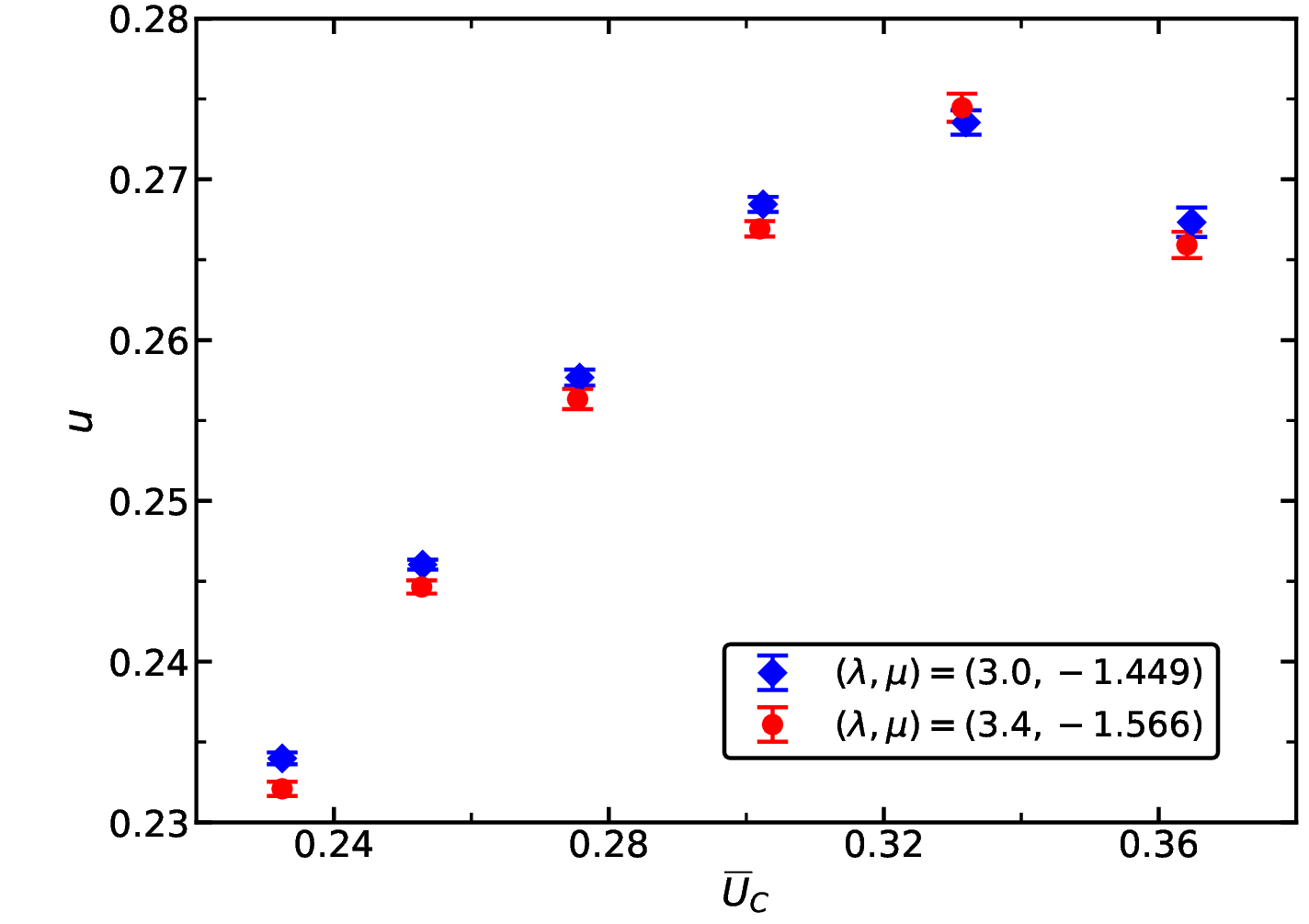}
\caption{\label{couplingplot1}
We plot estimates of $u$, eq.~(\ref{flowequation}), computed by using 
eq.~(\ref{diffdef}) and the data for $(\lambda,\mu)=(3.4,-1.566)$
and $(3.0,-1.449)$ as a function of $\overline{U}_C$. 
The data are
obtained for the pairs of lattice sizes $(L_1,L_2)=(12, 16)$, $(16, 24)$, 
..., $(64,96)$. Since $\overline{U}_C$ increases with $L$ in the given 
range, the leftmost points correspond to $(L_1,L_2)=(12, 16)$, while the
rightmost ones correspond to $(L_1,L_2)=(64, 96)$.
For a discussion see the text.
}
\end{center}
\end{figure}

To this end, we first compare the estimates
obtained from two different pairs of $(\lambda,\mu)$ that give approximately
the same values of $\overline{U}_C$ for the same lattice size $L$. 
The leading correction should differ between these two pairs.
In Fig. \ref{couplingplot1}  we plot our estimates of $u$ for 
$(\lambda,\mu)= (3.4,-1.566)$ and $(3.0,-1.449)$. For example, from the fit 
with the Ansatz~(\ref{R3master}), taking $m_{max}=9$, and
our largest set of data, we get for $L_{min}=24$ the 
estimates $w=-0.0038(5)$ and $0.0017(5)$, respectively. In particular
the difference between these two estimates of $w$ is very stable when 
varying the parameters of the fit. In both cases, estimates of $u$
obtained from the linear lattice sizes $L=12$, $16$, $24$, $32$, $48$, $64$,
and $96$ are shown. Note that $\overline{U}_C$ is monotonically increasing 
with the linear lattice size $L$ in the range plotted here. 
We find that the results obtained for $(L_1,L_2) = (12,16)$, which is leftmost
in the plot, and $(16,24)$ for the two pairs of $(\lambda,\mu)$
clearly differ by more than the statistical error. However, the difference 
is small compared with the value of $u$.

\begin{figure}
\begin{center}
\includegraphics[width=14.5cm]{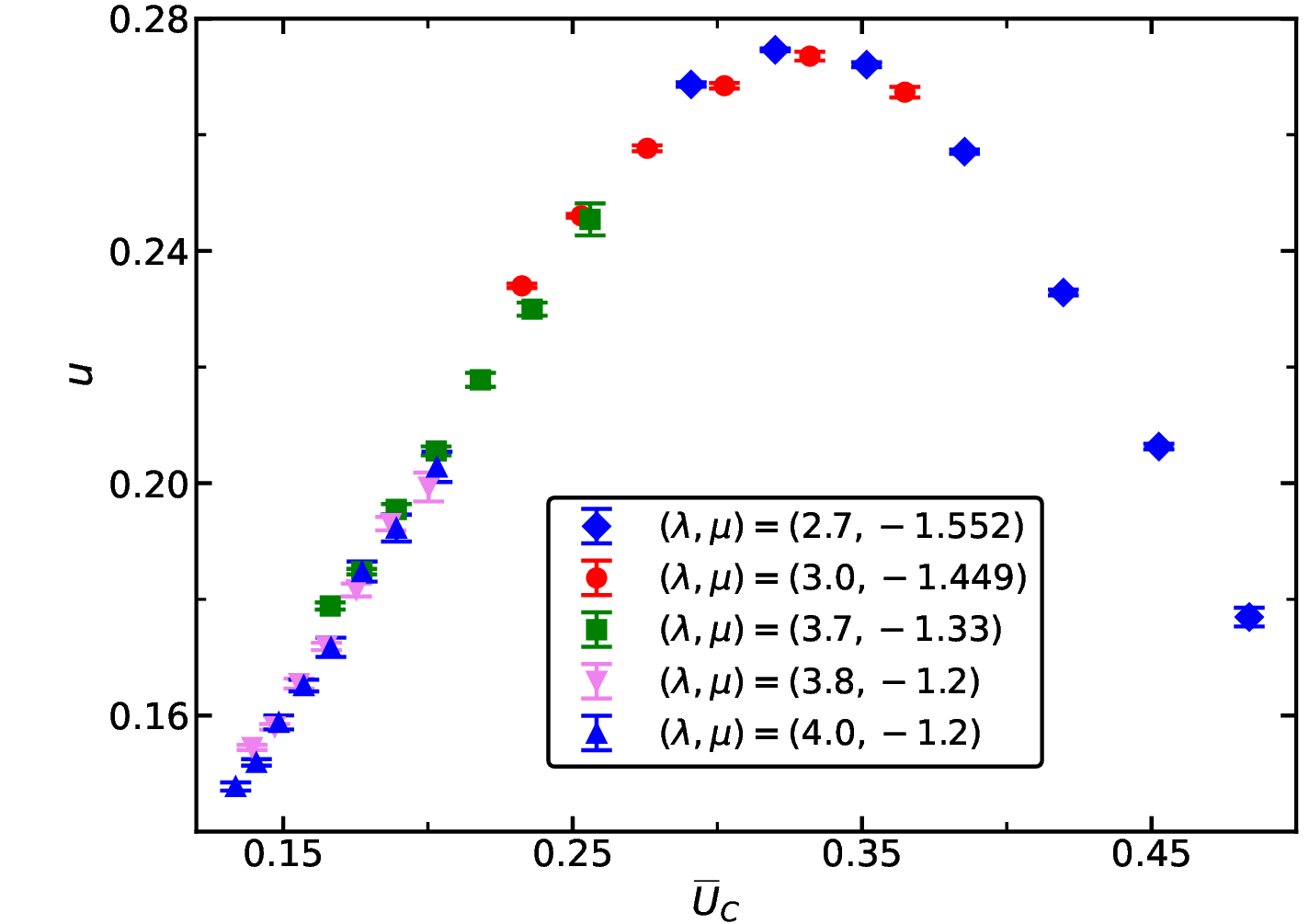}
\caption{\label{couplingplot0}
We plot estimates of $u$, eq.~(\ref{flowequation}),  computed by using 
eq.~(\ref{diffdef}) and the data for $(\lambda,\mu)= (2.7,-1.552)$,
$(3.0,-1.449)$, $(3.7,-1.3)$, $(3.8,-1.2)$, and $(4.0,-1.2)$,
as a function of $\overline{U}_C$.
For a discussion see the text.
}
\end{center}
\end{figure}

Next, in Fig. \ref{couplingplot0} we plot our estimates of $u$ 
computed by using eq.~(\ref{diffdef})
obtained for 5 different pairs of $(\lambda,\mu)$, which are approximately
on the line of slow flow. For small linear lattice sizes, we expect
that subleading corrections are the numerically dominant corrections. 
It is quite clear from the plot
that estimates obtained from $(L_1,L_2)=(12,16)$ are too large compared with 
the asymptotic value. Note that the data points for $(L_1,L_2)=(12,16)$ are 
the leftmost ones for each pair $(\lambda,\mu)$ which is considered.
For $(L_1,L_2)=(24,32)$ the inspection by eye does not
show such a deviation, suggesting that corrections to scaling are at 
most at the level of the statistical error at this point.

\begin{figure}
\begin{center}
\includegraphics[width=14.5cm]{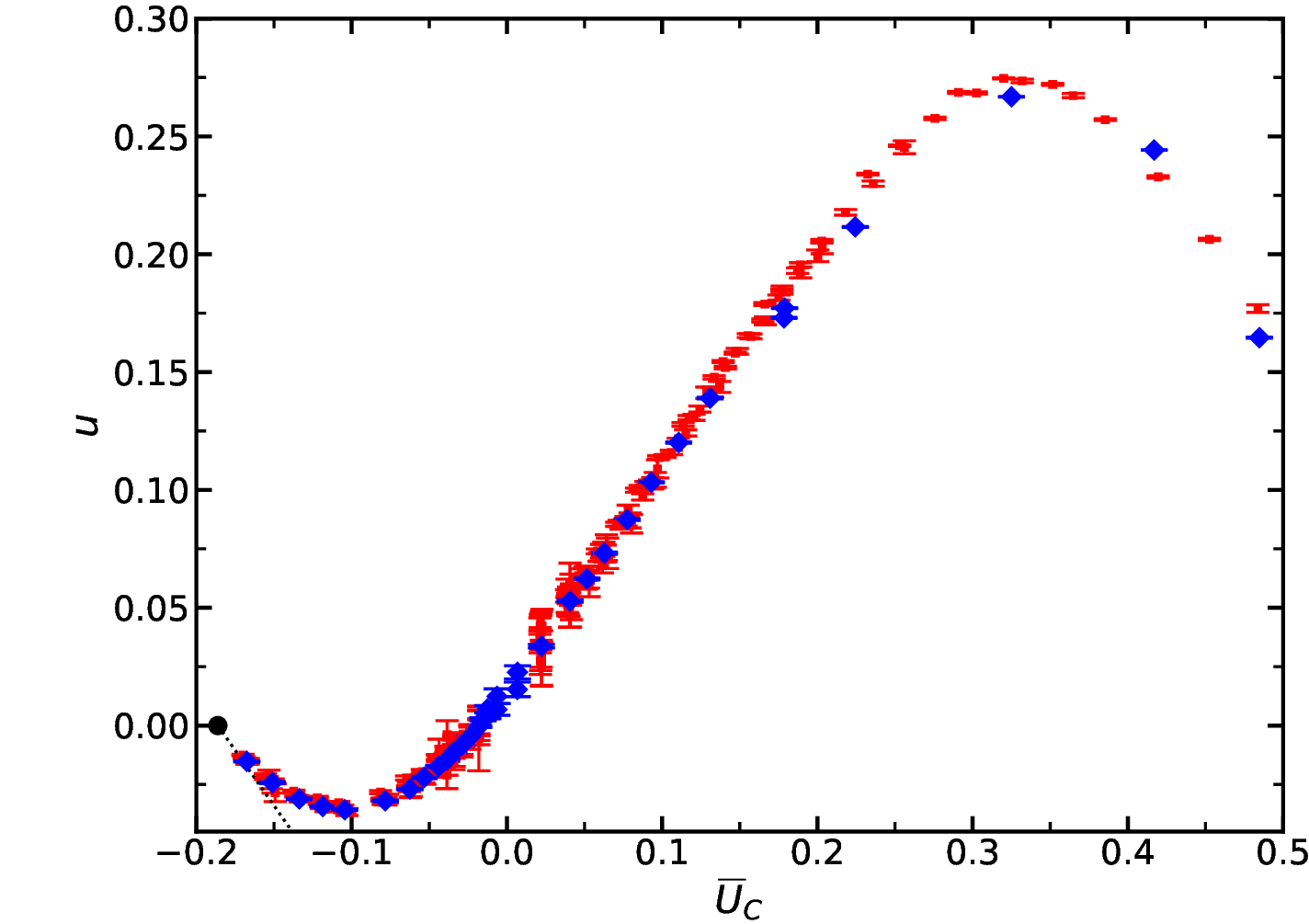}
\caption{\label{couplingplot}
We plot $u$, eq.~(\ref{flowequation}), obtained from a large set of 
$(\lambda,\mu)$ as a function of 
$\overline{U}_C$. Results obtained by using eq.~(\ref{mostsimple}) are
given as diamonds, while those obtained by using eq.~(\ref{diffdef})  are
given as squares. The filled circle gives the result for the decoupled
Ising system. The dotted line shows the behavior~(\ref{u_nextDI}) in the 
neighborhood of the decoupled Ising system.
Details are discussed in the text. 
}
\end{center}
\end{figure}

In Fig. \ref{couplingplot} we plot estimates of $u$ obtained by using 
eq.~(\ref{mostsimple}) with $L_{min}=24$ and eq.~(\ref{diffdef}) as a 
function of 
$\overline{U}_C$. Furthermore we give $u=0$ for the decoupled Ising
point at $\overline{U}_C=-0.186188(5)$ and the behavior of $u$ in the 
neighborhood of the decoupled Ising point. For  
$|u| \lessapprox 0.1$ the estimates obtained by using 
eq.~(\ref{mostsimple}) and eq.~(\ref{diffdef}) are consistent. For
$u>0.15$ we see clear differences.  For small $|u|$, the statistical 
error is quite large for eq.~(\ref{diffdef}). This is partially due to the 
fact that we simulated for more values of $L$ in the same range of lattice 
sizes than for larger $|u|$. 
The behavior for $\overline{U}_C \lessapprox -0.15$ seems to be 
consistent with the predictions for the decoupled Ising system and its
neighborhood. 

We analyze the numerical results by using the Ansatz
\begin{equation}
\label{uAnsatz}
 u = \sum_{i=0}^n a_i \overline{U}_C^i \;,
\end{equation}
where we have taken $n=3$, $4$ and $5$ in our fits. In the appendix 
\ref{appendixB}, we discuss the cases $n=1$ and $2$ analytically.
In our preliminary analysis, 
we experimented with various approaches. For example we combined data 
for $ |\mu| \le 0.6$  analyzed by using eq.~(\ref{mostsimple}) with 
data for $|\mu| > 0.6$ analyzed by using eq.~(\ref{diffdef}). 

In our final analysis, we use for simplicity only data with 
$|\mu| \ge 0.6$ analyzed by using eq.~(\ref{diffdef}). 
Note that
for $(\lambda,\mu)=(4.7,0.6)$ we have $\overline{U}_C(L) = -0.034532(8)$, ..., 
$-0.033963(22)$ for $L=12$, ..., $64$ and for $(\lambda,\mu)=(4.7,-0.6)$
we have $\overline{U}_C(L) = 0.047872(7)$, ..., $0.053152(27)$ 
for $L=12$,  ..., $64$.
In the fit, the covariances that are caused by the fact that the
numerical result for $\overline{U}_C(L)$ might appear in two differences,
one with a smaller and one with a larger lattice size, are taken into 
account. 
In order to estimate the effects of the truncation in eq.~(\ref{uAnsatz})
and corrections to scaling, we vary the maximal $|\overline{U}_C|$ and $|\mu|$
and minimal lattice size $L_{1,min}$ that is taken into
account. Our final results are based on four different fits.
In the first one, fit 1 in table 
\ref{resutable},  we included  $(\lambda,\mu)=(2.7,-1.552)$,
$(3.0,-1.449)$,
$(3.7,-1.33)$, $(3.8,-1.2)$, $(4.0,-1.2)$, $(4.2,-1.1)$, $(4.3,-1.0)$,
$(4.4,-0.9)$, $(4.5,-0.8)$, $(4.7,-0.7)$, $(4.7,-0.6)$, $(4.7,0.6)$,
$(4.7,0.7)$, $(4.5,0.8)$, $(4.3,1.0)$, $(4.0,1.2)$, $(3.4,1.5)$,
$(2.6,1.9)$, $(2.2,2.0)$, $(1.9,2.1)$, and $(1.65,2.2)$. 
In the case of $(\lambda,\mu)=(2.7,-1.552)$ we skipped the lattice 
sizes $L>64$, since the value of $\overline{U}_C$ is too large to be
be fitted with the Ansatz~(\ref{uAnsatz}). Furthermore,
we take $n=5$ in eq.~(\ref{uAnsatz}) and $L_{1,min}=48$. In the 
remaining fits, smaller ranges of $(\lambda,\mu)$ are used.
The values of $(\lambda,\mu)$ taken into account range from
$(3.7,-1.33)$  to $(1.65,2.2)$, $(4.0,-1.2)$ to $(2.2,2.0)$, and $(4.2,1.1)$
to $(2.6,1.9)$ for fits 2, 3, and 4, respectively. Furthermore, 
the fits 2, 3 and 4 are characterized by $L_{1,min}=32$, $24$, $24$ and 
$n=5$, $4$, $3$, eq.~(\ref{uAnsatz}), respectively. In the case of
fit 1, $\overline{U}_C$ ranges from $-0.15118(7)$ to $0.40551(4)$, while
for fit 4, it ranges from $-0.106242(13)$ to $0.13989(7)$. The
estimates of the parameters $a_i$ are summarized in table \ref{resutable}.
Based on these estimates, we compute
$\overline{U}_C^*$, which is the zero of $u$, numerically. Furthermore,
$\omega_2$ is given by minus the derivative of $\tilde u=\overline{U}_C u$
with respect to $\overline{U}_C$ at $\overline{U}_C^*$. The statistical 
errors of these derived quantities are obtained by error propagation as 
discussed in appendix \ref{appendixA}. It turns out that
the difference $Y_4-\omega_2$ has a much smaller statistical error than $Y_4=a_0$ 
and $\omega_2$, individually. This can be attributed to a strong statistical
correlation of $Y_4$ and $\omega_2$. The strong correlation is not 
surprising, since 
for $a_{i\ge2}=0$ we get $Y_4=\omega_2$, see appendix \ref{appendixB}.
The final results given in table \ref{resutable} and their error bars
are chosen such that the results of the four fits discussed above
are covered. For a discussion of the error analysis see 
appendix \ref{appendixA}. 

\begin{table}
\caption{\sl \label{resutable}
Fitting numerical estimates of $u$, eq.~(\ref{diffdef}), by using the 
Ansatz~(\ref{uAnsatz}). The fits, which
are labelled by 1, 2, 3, and 4 are discussed in the text. In addition to
the estimates given in the table, we obtain $a_5=-18.1(2.5)$ and 
$42.7(9.3)$ for the fits 1 and 2, respectively. We get 
$\chi^2/$DOF$=0.965$, $0.970$, $0.934$, and $0.968$
corresponding to $p=0.524$, $0.538$, $0.623$, and $0.542$
for fits 1, 2, 3, and 4, respectively. In the case of the fits 1, 2, 3, and 4
we give statistical errors, while for the final results systematic errors
are taken into account, as discussed in appendix \ref{appendixA}. }
\begin{center}
\begin{tabular}{ccccccccc}
\hline
fit &$a_0$&$a_1$&$a_2$&$a_3$&$a_4$&$\overline{U}_C^*$ &$\omega_2$ & $Y_4-\omega_2$\\
\hline
 1 & 0.01441(66) & 0.8136(72) & 2.004(55) & $-$8.7(4) & 12.9(1.9)
   & $-$0.0186(8) &0.01360(59) & 0.00081(6) \\
 2 & 0.01348(38) & 0.8296(34) & 2.398(69) & $-$10.5(3) &$-$1.4(2.6)
   & $-$0.0172(4) & 0.01267(31) & 0.00081(3) \\
 3 & 0.01448(21) & 0.8367(25) & 2.193(41) & $-$10.5(2) & 10.0(1.9)
   & $-$0.0183(3)  & 0.01362(20) &  0.00086(2) \\
 4 & 0.01401(18)  & 0.8358(28) & 2.357(23) & $-$10.4(3)  &     
   & $-$0.0177(2) & 0.01315(16) & 0.00086(2) \\
\hline 
final & 0.0141(10) & 0.823(17) &  2.21(26)  & $-$9.6(1.3) & 
 & $-$0.0181(14) & 0.0133(9) & 0.00081(7) \\
\hline
\end{tabular}
\end{center}
\end{table}

As a final check, we repeated the analysis, replacing 
$(\lambda,\mu)=(3.7,-1.33)$,
$(3.8,-1.2)$, $(4.0,-1.2)$, $(4.2,-1.1)$, and $(4.3,-1.0)$  by
$(\lambda,\mu)=(3.0,-1.663)$, $(3.4,-1.8)$, $(3.7,-1.5)$, $(4.0,-1.2)$,
and $(4.5,-1.0)$. For these values of $(\lambda,\mu)$ the amplitude $|w|$ 
of corrections is larger than for the replaced ones. The results do not
change significantly.

\subsection{matching}
For two pairs of parameters $(\lambda_1,\mu_1)$ and $(\lambda_2,\mu_2)$ we 
determine a scale factor $c_{1,2}$ by requiring that
\begin{equation}
\label{matching_condition}
\overline{U}_{C,1}(L) =  \overline{U}_{C,2}(c_{1,2} L) \;, 
\end{equation}
where the second subscript indicates the pair of parameters.
This is solved numerically for each linear lattice size that we simulated 
for parameter pair one.
In a first step, for $\mu < 0$, 
we determine two lattice sizes $L_{1}$ and $L_{2}$ for the 
second parameter pair such that $L_2$ is the smallest linear lattice size 
simulated such that $\overline{U}_{C,1}(L) \le \overline{U}_{C,2}(L_2)$
and $L_1$ the largest such that 
$\overline{U}_{C,1}(L) \ge \overline{U}_{C,2}(L_1)$.
If such a pair of lattice sizes exists, we interpolate 
$\overline{U}_{C,2}$ linearly 
in the logarithm of the linear lattice of the second parameter pair.
As an example, 
in table \ref{matching1}, we give the results of the matching for
$(\lambda_1,\mu_1)=(2.333,-1.764)$ and $(\lambda_2,\mu_2)=(2.7,-1.552)$.
Note that for $(\lambda,\mu)=(2.333,-1.764)$
we find below in Sec. \ref{FirstOrder} that $\xi_{high} = 24.70(2)$ at the
transition temperature. Furthermore, for $(\lambda,\mu)=(2.333,-1.764)$
we reach linear lattice sizes, where $\overline{U}_{C}(L)$ 
becomes non-monotonic. We get $\overline{U}_{C} = 0.39042(10)$, 
$0.41910(4)$, $0.45781(5)$, $0.48359(5)$, $0.51161(6)$, $0.50655(12)$
for $L=12$, $16$, $24$, $32$, $48$, and $64$.  The linear lattice sizes 
given in table \ref{matching1}, are still in the range, where 
$\overline{U}_{C}(L)$ monotonically increases with the linear 
lattice size $L$. 

We find that $c_{1,2}$ changes only little  with
increasing $L$. It seems plausible that for $L=32$ systematic errors
are at most of the same size as the statistical error given 
in table \ref{matching1}.

\begin{table}
\caption{\sl \label{matching1}
Results for the scale factor $c$ for the matching between 
$(\lambda,\mu)=(2.333,-1.764)$ and $(2.7,-1.552)$. $L$ is the
linear lattice size for $(\lambda,\mu)=(2.333,-1.764)$ and 
$c_{1,2}$ gives the ratio with the matching lattice 
for $(\lambda,\mu)=(2.7,-1.552)$ as defined by 
eq.~(\ref{matching_condition}). For a discussion see the text.
}
\begin{center}
\begin{tabular}{cc}
\hline
$L$ &  $c_{1,2}$  \\
\hline
12 & 3.4351(35) \\
16 & 3.4478(18) \\
24 & 3.4558(23) \\
32 & 3.4523(44) \\
\hline
\end{tabular}
\end{center}
\end{table}

To check, whether $c_{1,2}$ indeed gives the ratio of the correlation 
length $\xi_{high}$ at $(\lambda_1,\mu_1)$ and $(\lambda_2,\mu_2)$, 
 we performed the matching for 
$(\lambda_1,\mu_1)=(2.0,-1.85)$ and $(\lambda_2,\mu_2)=(2.333,-1.764)$.  
Here we get
$c_{1,2}=2.0165(30)$, $2.0168(14)$, and $2.0139(15)$ for 
$L=12$, $16$, and $20$, respectively.   This can be compared with the
ratio $24.70(2)/12.135(6)=  2.0354(19)$   
of the correlation length in the high temperature phase computed below
in Sec. \ref{FirstOrder}.  The correction to the expected scaling is
small.

We continued this matching for pairs $(\lambda_1,\mu_1)$ and
$(\lambda_2,\mu_2)$ that are approximately
on the line of slow flow. In table \ref{matching2} we report our final
results for the matching factor $c_{1,2}$. 
The error bar includes a rough estimate
of the systematic error, obtained from the variation of $c_{1,2}$ 
with increasing 
$L$. Based on our simulations, we can not proceed to $\mu > -1$, since 
we have no pairs of $(\lambda,\mu)$ at hand that have overlapping ranges 
of $\overline{U}_C$. In the third column, we give an estimate
of the correlation length in the high temperature phase at the transition
temperature. We start from the direct estimate obtained for 
$(\lambda,\mu)=(2.333,-1.764)$ in Sec. \ref{FirstOrder} below. 
Then we multiply up
the values for $c_{1,2}$. The error bar is simply computed by adding up the 
error due to the previous estimate of $\xi_{high}$ and the one due to the 
uncertainty
of the current value of $c_{1,2}$. This is done since we do not know how 
the errors are correlated.

\begin{table}
\caption{\sl \label{matching2}
Results for the scale factor $c_{1,2}$ for a sequence of pairs
of $(\lambda,\mu)$. The pairs $(\lambda,\mu)$ are given in the first column.
We match subsequent pairs of $(\lambda,\mu)$. The estimates of $c_{1,2}$ 
given in the second column refer to $(\lambda_1,\mu_1)$ given
one row above, and $(\lambda_2,\mu_2)$ given in the same row. The estimate
of $\xi_{high}$ given in the third column is obtained by multiplying up 
the values for $c_{1,2}$, starting from $\xi_{high}=24.70(2)$ for the correlation
length in the high temperature phase at the transition temperature at
$(\lambda,\mu)=(2.333,-1.764)$. The errors for the correlation 
length are added up. For a discussion see the text.
}
\begin{center}
\begin{tabular}{ccc}
\hline
$(\lambda,\mu)$   &    $c_{1,2}$    &  $ \xi_{high} $ \\
\hline
$(2.333,-1.764)$  &           & 24.70(2) \\
$(2.7,-1.552)$  & 3.455(7)  & 85.34(24)  \\
$(3.0,-1.449)$  & 2.463(10)  & 210.2(1.4)  \\
$(3.7,-1.33)$   & 5.31(2)  & 1116.(12.)  \\
$(3.8,-1.2) $   & 2.99(2)  & 3337.(58.) \\
$(4.0,-1.2)$    & 1.33(1)  & 4438.(110.) \\
$(4.2,-1.1)$    & 3.35(4)  & 14870.(550.)  \\
$(4.3,-1.0)$    & 3.6(1)  & 53500.(3500.)  \\
\hline
\end{tabular}
\end{center}
\end{table}

Going to $\mu>-1$ we evaluate the RG-flow by using eq.~(\ref{flowequation}).
Here we abstain for simplicity from propagating the errors of the coefficients.
Instead we run the integration with the results for the coefficients 
$a_i$,  eq.~(\ref{uAnsatz}), of four different fits.
The spread of the results serves as rough estimate of the error.
Let us first check the consistency with the results given in table
\ref{matching2}. Let us consider $(\lambda_1,\mu_1)=(3.8,-1.2)$ and 
$(\lambda_2,\mu_2)=(4.2,-1.1)$
as example, where $\overline{U}_C= 0.179833(39)$ and  $0.139891(65)$
for $L=64$, respectively. Running eq.~(\ref{flowequation}) with the 
coefficients obtained from four different fits, we arrive at the estimate 
of the scale factor $c_{1,2}=4.70(12)$, which can be compared with 
$c_{1,2}=1.33(1) \times 3.35(4) = 4.46(9)$ taken from table \ref{matching2}.
Next we computed the scale factor $c_{1,2}$ 
between $(\lambda_1,\mu_1)=(3.8,-1.2)$
and $(\lambda_2,\mu_2)=(4.5,-0.8)$, $(4.7,-0.6)$, and $(5.0,-0.3)$ by 
using eq.~(\ref{uAnsatz}) as examples.
We get $c_{1,2}=562.(26.)$, $152000.(17000.)$, and $3.9(1.0) \times 10^{13}$,
respectively. Hence the correlation length in the high temperature phase
at the transition temperature should be $\xi_{high}=1960000.(90000.)$, 
$5.5(6) \times 10^8$, and $1.3(3) \times 10^{17}$, respectively.
Note that the estimates of the error are only rough ones. Still the order
of magnitude of $\xi_{high}$ should be correct. It is apparent that the 
range of parameters, where the first order transition is very weak,
is large.

\section{effective exponent of the correlation length}
\label{nueff}
In ref. \cite{AharonyNeu} the authors
suggest that for a weak first order transition, for a large range of 
reduced temperatures, the behavior
of the correlation length is similar to that at a second order phase 
transition, where however the exponent $\nu$ of the $O(3)$-invariant 
Heisenberg universality class is replaced by an effective one that 
depends weakly on the reduced temperature.  Here we analyze the 
finite size scaling behavior of the slope of dimensionless quantities and the 
behavior of the infinite volume correlation length in the high temperature 
phase.

\subsection{Finite size scaling}
We analyze the slopes $S_i$ of dimensionless quantities 
$\tilde R_i = R_i - \sum_{j=2}^m c_{i,j} U_C^j$
at $\beta_f$, where $S_i=\partial \tilde R_i/\partial \beta$. 
The idea is that $\tilde R_i$ stays approximately constant 
with increasing $L$ at the transition temperature, and that this hopefully 
also improves the  behavior of the slope $S_i$. We redo the
analysis of section VII D of ref. \cite{myCubic} with new data added.
Note that in ref. \cite{myCubic} we have used by mistake the wrong sign for 
the improvement term
$\sum_{j=2}^m c_{i,j} \overline{U}_C^j$. Here we compare final results 
obtained by using different choices of $c_{i,j}$.

\begin{figure}
\begin{center}
\includegraphics[width=14.5cm]{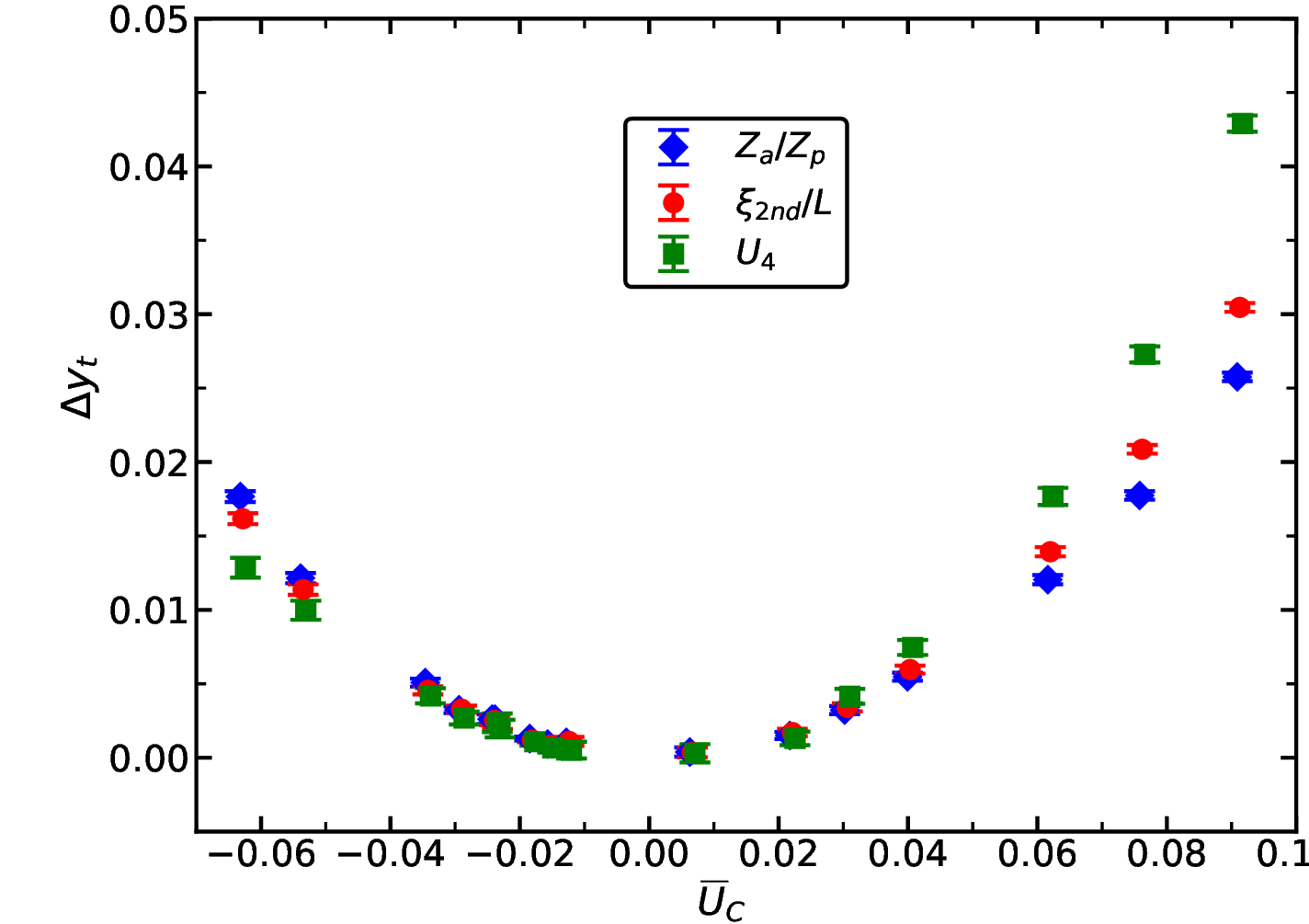}
\caption{\label{nueffnew}
We plot $\Delta y_{t}$, eq.~(\ref{rAnsatz}), obtained for a large set of
$(\lambda,\mu)$ as a function of $\overline{U}_C$.
Details are discussed in the text.
}
\end{center}
\end{figure}

Since we are interested in the difference compared with the Heisenberg 
universality class, we analyze ratios
\begin{equation}
r_{S,i}[(\lambda,\mu), (\lambda_0,0), L] = 
\frac{S_{\lambda,\mu,i}(L)}{S_{\lambda_0,\mu=0,i}(L)} \; ,
\end{equation}
where $i$ indicates which dimensionless quantity is taken and 
$\lambda_0=5.2$ or $5.0$. Note that $\lambda^*=5.12(5)$ for $\mu=0$,
see section \ref{slowline}.  
We expect that subleading corrections approximately cancel. 
Therefore we analyze the ratio with the simple Ansatz
\begin{equation}
\label{rAnsatz}
r_{S,i}[(\lambda,\mu), (\lambda_0,0), L] = a L^{\Delta y_{t}} \; .
\end{equation}
We performed fits for a number of values of $(\lambda,\mu)$ using a minimal 
lattice size $L_{min}=16$ or $24$ that is taken into account.
In Fig. \ref{nueffnew} we plot $\Delta y_{t}$ obtained by using  $L_{min}=24$
and $\lambda_0=5.2$ as a function of $\overline{U}_C$. The values of $c_{i,j}$
are taken from table \ref{coeff_c} and $m=4$. Here $\beta_f$ is obtained from 
fixing $Z_a/Z_p-\sum_{j=2}^m c_{i,j} U_C^j=0.19477$ using the values of
$c_{i,j}$ given in table \ref{coeff_c}.
As argument of $\Delta y_{t}$ we take 
$[\overline{U}_C(L_{max}) + \overline{U}_C(L_{min})]/2$, where $L_{max}$
and $L_{min}$ are the largest and the smallest lattice size taken into 
account in the fit.

We have analyzed the estimates of $\Delta y_{t}$ by using the Ans\"atze 
\begin{equation}
\label{Delta_Ansatz1}
\Delta y_{t}  = b \overline{U}_C^2 + c \overline{U}_C^3
\end{equation}
and 
\begin{equation}
\label{Delta_Ansatz2}
\Delta y_{t}  = b \overline{U}_C^2 + c \overline{U}_C^3 + d \overline{U}_C^4 
\end{equation}
for the three different dimensionless quantities with different choices
for $\sum_{j=2}^m c_{i,j} \overline{U}_C^j$. 

For $c_{i,j}=0$, already the estimate of $b$ depends on the dimensionless
quantity that is considered. For example from a fit with $L_{min}=16$ and 
$\lambda_0=5.2$ we get $b=3.94(3)$, $4.21(3)$, and $3.55(5)$ for $Z_a/Z_p$, 
$\xi_{2nd}/L$, and $U_4$, respectively. In all three cases we used the 
Ansatz~(\ref{Delta_Ansatz1}) and data for
$-0.06 \lessapprox \overline{U}_C  \lessapprox 0.06$.

We added successively higher orders of $\overline{U}_C$ to define $\tilde R_i$.
We could not identify a clean convergence pattern for the coefficients of 
eqs.~(\ref{Delta_Ansatz1},\ref{Delta_Ansatz2}).

For the choice of $c_{i,j}$ used for the data given in Fig. \ref{nueffnew},
we find that the results for $b$ are more or less the same for the three 
different dimensionless quantities. We quote 
\begin{equation}
\label{beq}
 b = 3.8(2) 
\end{equation}
as our final result. For $c$, the results depend clearly on the 
dimensionless quantity that is considered. Certainly deeper theoretical insight 
is needed to decide whether a unique effective exponent $\nu_{eff}$ can be 
obtained from finite size scaling.

Note that at the cubic fixed point, for finite $m$ and any choice of 
$c_{i,j}$, one should get in the limit $L\rightarrow \infty$ a unique value for 
$\Delta y_t$, not depending on the choice of the dimensionless quantity.
Indeed, analyzing various choices $\sum_{j=2}^m c_{i,j} \overline{U}_C^j$, 
we get similar numerical estimates for $\Delta y_t$ at the cubic fixed point
for $Z_a/Z_p$, $\xi_{2nd}/L$, and $U_4$.
In particular we confirm the numerical results of Sec. VII D of 
ref. \cite{myCubic}. In particular, plugging in the numerical estimate of 
$\overline{U}_C^*$ obtained in section \ref{UCflow} above we get
\begin{equation}
y_{t,cubic} - y_{t,O(3)} = 3.8(2) \times [-0.0181(14)]^2 = 0.00124(12) \;,
\end{equation}
where we have ignored $O(\overline{U}_C^{* \; 3})$ contributions and have
added up the errors due to $b$, eq.~(\ref{beq}), and $\overline{U}_C^*$.

\subsection{Correlation length in the high temperature phase}
Here we have simulated the model for the two selected values 
$(\lambda,\mu)=(4.5,-0.8)$ and $(3.8,-1.2)$
in the high temperature phase, where the correlation length can be
determined very accurately by using the improved estimator of the 
correlation function that comes with the single cluster algorithm
\cite{Wolff}. 
For comparison we study $\lambda=5.0$ and $5.2$ at $\mu=0$.

As estimate of the correlation length we take the effective correlation length
\begin{equation}
\label{xieff}
\xi_{eff}(t) = -\ln(G(t+1)/G(t)) \;,
\end{equation}
where $G(t) = \langle \vec{S}(0) \cdot \vec{S}(t)  \rangle$  and
\begin{equation}
\vec{S}(x_0) = \sum_{x_1,x_2} \vec{\phi}_{x_0,x_1,x_2} \;.
\end{equation}
Computing $G(t)$, we summed over all translations and all three 
directions on the lattice.

In the numerical analysis we improved eq.~(\ref{xieff}) by 
taking into account periodic boundary conditions 
\begin{equation}
\label{xieff_periodic}
G(\tau) =c [ \exp(-\tau/\xi_{eff}) + \exp(-(L-\tau)/\xi_{eff}) ] \;.
\end{equation}
Eq.~(\ref{xieff_periodic}), for $\tau=t$ and $\tau=t+1$, is solved for
$\xi_{eff}$ numerically.
It turns out that $\xi_{eff}(t)$ is rapidly converging with increasing 
distance $t$. As our final estimate we take  $\xi_{eff}(t)$ at
$t=2 \xi_{eff}(t)$, self-consistently.  We take a linear lattice size
$L \approx 20 \xi_{eff}$.  We checked that for this lattice size finite 
size effects are clearly negligible.

We performed simulations for a range of $\beta$ such that $\xi \approx 2$
for the smallest value of $\beta$ and $\xi \approx 10$ for the largest.
We simulated at  26, 29, 34, and 36 different values of $\beta$ for
$(\lambda,\mu)=(3.8,-1.2)$, $(4.5,-0.8)$, $(5.0,0)$, and $(5.2,0)$, 
respectively.
We performed at least $10^5$ update cycles for each simulation. 
The update cycle consist of local Metropolis updates and single cluster
updates. We performed roughly as many single cluster updates, such that,
on average, the volume of the lattice is covered.

Assuming that the models are improved, we fitted our data with the simple
Ansatz
\begin{equation}
\label{xiAnsatz}
\xi=a t^{-\nu_{eff}} \; (1 + b t)  \;,
\end{equation}
where we have included leading analytic corrections. Our definition of the 
reduced temperature is $t=\beta_t-\beta$. In a way, along with the
range of $\beta$ that is taken into account in the fit, this defines an 
effective value of the correlation length exponent $\nu$. 
We took the estimate of $\beta_t$ from the finite size scaling analysis 
discussed above. The parameters of the fit are $a$, $b$, and $\nu_{eff}$.

Fitting all our data for $(\lambda,\mu)=(5.2,0)$ we get $\chi^2/$DOF$=1.237$
corresponding to $p=0.164$ and $\nu_{eff}=0.71045(8)$, which is slightly
too small compared with $\nu = 0.71164(10)$ \cite{myIco} or 
$\nu = 0.71169(30)$ \cite{CB_O3}. 
Discarding small values of $\beta$, the fit improves and the value of 
$\nu $ increases. For example taking $\beta=0.65$ with $\xi=3.25846(24)$
as minimal value, we get $\chi^2/$DOF$=1.005$ corresponding to $p=0.455$
and $\nu_{eff}=0.71093(22)$. The small deviation of $\nu_{eff}$ from the 
estimates of refs. \cite{myIco,CB_O3} can be attributed to corrections
not taken into account in the Ansatz~(\ref{xiAnsatz}). Analyzing the data
for $(\lambda,\mu)=(5.0,0)$ we get $\chi^2/$DOF$=0.999$ 
corresponding to $p=0.467$ and $\nu_{eff}=0.71024(8)$.  Also in this case, 
$\chi^2/$DOF decreases, when discarding small values of $\beta$ and the 
value of $\nu_{eff}$ slightly increases. For example for the minimal 
value $\beta=0.65$ with $\xi=3.27436(28)$ we get $\chi^2/$DOF$=0.680$, 
$p=0.870$ and $\nu_{eff}=0.71095(24)$. 

In summary: Using the simple Ansatz~(\ref{xiAnsatz}) taking data for 
$\xi \approx 3.3$ up to $\xi \approx 10$ we obtain an estimate of 
$\nu$ that deviates from the most accurate values for the Heisenberg
universality class, given in the literature, in the fourth digit.

Now let us turn to the data for $\mu < 0$. 
For $(\lambda,\mu)=(4.5,-0.8)$ we get $\chi^2/$DOF$=1.550$, 
$p=0.0365$ taking into account all data. We get $\nu_{eff}=0.70235(9)$. 
The quality of the fit does not improve discarding data. We note that 
our numerical estimates of $\xi$ are very accurate and less accurate 
data might result in an acceptable fit. The estimate of $\nu_{eff}$ is 
clearly smaller than those obtained for $\mu=0$. The deviation is about
$1 \%$. 

Finally we analyzed our data for $(\lambda,\mu)=(3.8,-1.2)$. 
Fitting all data
we get $\chi^2/$DOF$=5.054$,  $p=0.000$, and $\nu_{eff}=0.68382(8)$.
Here, discarding data, keeping $\beta=0.64$ , $\xi=3.61272(27)$ as smallest
value of $\beta$ we get $\chi^2/$DOF$=1.309$, $p=0.199$ and 
$\nu_{eff}=0.68176(27)$, which is clearly smaller than the $O(3)$ invariant
value. 
Fitting all data up to $\beta=0.648$, $\xi=4.50359(34)$, we get 
$\chi^2/$DOF$=0.692$, $p=0.733$  and $\nu_{eff}=0.68519(28)$.  We notice that
the value of $\nu_{eff}$ decreases, decreasing the reduced temperature $t$. 

In order to compare the result obtained here with that obtained from finite
size scaling, we take $\overline{U}_C$ as defined in Sec. \ref{UCflow}
for $L=8$ at $(\lambda,\mu)=(4.5,-0.8)$ and $(3.8,-1.2)$, where 
$L=8$ is chosen to have a rough match with the
correlation lengths up to $\xi \approx 10$ that are considered here.
We estimate $\overline{U}_C \approx 0.0678$ and $0.1177$, respectively.
Hence we get $y_{t,eff} \approx 1.4052 + 3.8 \overline{U}_C^2$ 
$\approx 1.4227$ and $1.4578$, corresponding to $\nu_{eff} \approx 0.7029$
and $0.6859$, respectively.  These numbers are in reasonable agreement
with the results obtained from the correlation length in the high 
temperature phase.

\section{First order phase transition}
\label{FirstOrder}
Here we discuss our simulations for values of $(\lambda,\mu)$, where
the first order transition is sufficiently strong such that it can be detected
directly in the analysis of the data generated in the simulation.
We performed extensive preliminary simulations to get an idea of the 
range of $(\lambda,\mu)$, where this is the case.
A first indication of a first order transition is the appearance of 
metastabilities in standard simulations. Furthermore,
it is useful to study the histograms of various observables. At first 
order transitions, double peak structures appear. These double peaks 
become sharper and clearer separated 
as the linear lattice size increases. The separation 
of the peaks is accompanied by a rapid increase of the autocorrelation 
time with increasing lattice size, when using standard algorithms.

Below, we briefly discuss our implementation of the multicanonical method 
\cite{BeNe91,BeNe92} that at least mitigates the problem of the increasing 
autocorrelation time. For more detailed discussions and alternatives to 
the multicanonical method see for example refs. 
\cite{WaLa01,WaLa01E,EaDe05,Ja98,Ja02}.
Then we discuss our numerical results for the transition temperatures, 
the interface tension, the latent heat and the correlation length in the
high temperature phase at the transition.  The theoretical basis for the 
finite size scaling analysis of first order phase transitions is provided 
by refs. \cite{BoKo90,BoJa92}. 

\subsection{multicanonical method}
In order to perform simulations for lattices with $L \gg \xi_{high}$ 
at the transition temperature, we employed the multicanonical method 
\cite{BeNe91,BeNe92}.  In standard simulations, using a local algorithm, 
configurations can be changed only in small steps. Hence going from 
the disordered to an ordered phase and vice versa, the Markov 
chain has to pass configurations, where both phases are present, separated
by interfaces. These configurations are highly suppressed and their weight
is decreasing exponentially with the area of the interfaces. Therefore, 
in the simulation these configurations are rarely visited and hence 
tunneling times between the phases become larger and larger as the 
lattice size increases.

The basic idea of the multicanonical method is to simulate 
a modified distribution such that configurations that contain two phases
have an enhanced probability compared with the Boltzmann distribution.
Configurations $\{\vec{\phi}\}$ are generated with a probability distribution
\begin{equation}
P(\{\vec{\phi}\}) = \frac{1}{\sum_{\{\vec{\phi}\}} \exp\left(-H[\{\vec{\phi}\}]\right) 
W(X[\{\vec{\phi}\}]) } \;
\exp\left(-H[\{\vec{\phi}\}]\right) W(X[\{\vec{\phi}\}])  \;,
\end{equation}
where $W(X[\{\vec{\phi}\}])$ is a real positive number and $X[\{\vec{\phi}\}]$ is an 
estimator of an observable. In our simulations we took the 
energy, eq.~(\ref{energy}), for this purpose.

Using the multicanonical method the problem of the increasing 
tunneling time can be drastically reduced but not completely 
eliminated. For a discussion see for example ref. \cite{Bauer10}.

The expectation value of an estimator $A[\{\vec{\phi}\}]$ with respect to the 
Boltzmann distribution is given by
\begin{equation}
 \langle A \rangle \approx \frac{\sum_i W^{-1}(X[\{\vec{\phi}\}_i]) A[\{\vec{\phi}\}_i]}
  {\sum_i W^{-1}(X[\{\vec{\phi}\}_i]) }  \;,
\end{equation}
where we sum over the configurations that are generated after 
equilibration. 

The function $W(X)$ should be constructed such that the histogram 
becomes essentially flat between the maxima of the Boltzmann distribution.
We construct $W(X)$ as a piecewise constant function:
\begin{equation}
W(X)  = \left\{   
\begin{array}{@{\,}c@{}ll}
1 \;\;   &\mbox{for}&  X<X_0 \\
                          w_i  \;\; &\mbox{for}&  X_0+ i \Delta \le X < 
                                (i+1) \Delta\\
                          1  \;\;   &\mbox{for}& X_1 < X \;,
\end{array}\right.
\end{equation}
where $i \in \{0, 1, ..., M-1 \}$ and $\Delta=(X_1-X_0)/M$.  $X_0$ and $X_1$ 
roughly give the position 
of the peaks in the histogram. In our simulations $10 \le M \le 600$.
The weights $w_i$ are computed from the histogram. They can 
be iteratively improved by using more and more accurate data 
for the histogram. For lattice sizes that are not too large, 
one gets a few tunnelings between the phases by simulating with the 
Boltzmann distribution and one can use these simulations as starting 
point for the iterative determination of $W(X)$. 

In case one has a reasonable Ansatz for the histogram of $X$ as a function
of the linear lattice size $L$, one might increment $L$ in small steps.
A first guess for $W(X)$ might be obtained by extrapolating the results
obtained for the lattice sizes simulated before.

Here we did not succeed with such a strategy.  Instead,  we proceed
without using the knowledge obtained from the simulation of smaller
lattice sizes:
We started with two simulations taking $W(X)=1$ for all $X$. These
simulations are started with configurations that are in the domain of
the disordered and the ordered phase, respectively.
For the disordered phase we take
\begin{equation}
 \phi_{x,i} = \mbox{rand} -0.5
\end{equation}
for all sites $x$ and components $i$, where $\mbox{rand}$ is a uniformly 
distributed random number in the interval $[0,1)$.
In the case of the ordered phase we take
\begin{equation}
\label{order1}
 \phi_{x,0} = \Phi_0 + \mbox{rand} -0.5
\end{equation}
and 
\begin{equation}
\label{order2}
 \phi_{x,i} = \mbox{rand} -0.5
\end{equation}
for $i>0$, where $\Phi_0$ is a rough approximation of the expectation
value of the field in the ordered phase.

For sufficiently large lattice sizes $L$, the probability that the 
simulation switches the phase during the simulation is virtually 
vanishing.
We compute the histograms of $X$ for these two simulations.  
We chose $X_0$ as the position of the maximum of the histogram
of the disordered simulation and $X_1$ as the position of the 
maximum of the histogram of the ordered simulation. Typically we
get reasonable statistics only up to $X_0+\epsilon_0$ and 
down to $X_1-\epsilon_1$. In the middle, there is a gap without any 
configuration generated. We compute $W(X)$ up to $X_0+\epsilon_0$ and
and down to $X_1-\epsilon_1$ straight forwardly from the histogram. 
The gap between $X_0+\epsilon_0$ and $X_1-\epsilon_1$ is filled
by linear interpolation.  We also experimented with guessing somewhat
larger values of $W(X)$ in the gap to speed up the convergence.
We iterated this step until the gap has closed. Then we proceeded
as above.

We performed the simulations using a hybrid of local Metropolis, local 
overrelaxation and wall cluster \cite{HaPiVi} updates.
The weight $W$ is integrated in the 
accept/reject step of the local algorithms in the straight forward 
way. In the case of the wall cluster algorithm, the cluster is 
constructed following the same rules as for the plain Boltzmann 
distribution. The update of the wall cluster is viewed as a proposal
of a Metropolis step, where the accept/reject step takes into account
the change of $W$ caused by the wall cluster update:
\begin{equation}
P_{acc} = \mbox{min}[1,W(X[\{\vec{\phi}\}'])/W(X[\{\vec{\phi}\}])]  \;,
\end{equation}
where $\{\vec{\phi}\}'$ is the configuration that results from the wall cluster
update of $\{\vec{\phi}\}$.

\subsection{Simulations at the first order transition}
Based on our preliminary studies we focussed on simulations for the five pairs 
of parameters:  $(\lambda,\mu)=(1.24,-2.3)$, $(1.675,-1.95)$, $(2.0,-1.85)$, 
$(3,-2.5)$, and 
$(2.333,-1.764)$. These values were selected such that the correlation
length in the high temperature phase, at the transition temperature is about 
$\xi_{high} \approx 2$, $6$, $12$, $12$, and $24$, respectively.
The smaller $\xi_{high}$, the stronger is the first order transition. 

First, for all pairs of parameters,
we performed simulations with the program used in ref. \cite{myCubic}, 
generating configurations following the Boltzmann distribution.
It can be used as long as the tunneling times between the phases are
not too large. We performed such 
simulations using the linear lattice size $L=8$ for 
$(\lambda,\mu)=(1.24,-2.3)$, 
$L=12$, $16$, and $24$  for 
$(\lambda,\mu)=(1.675,-1.95)$, 
$L=12$, $16$, $20$, $24$, $32$, $40$, and $48$  
for $(\lambda,\mu)=(2,-1.85)$, $L=12$, $16$, $20$, $24$, and $32$ for 
$(3,-2.5)$, and  $L=12$, $16$, $20$, $24$, $32$, $40$,
$48$, and $64$ for $(\lambda,\mu)=(2.333,-1.764)$.
We extracted a preliminary estimate of the transition temperature by 
requiring that $Z_a/Z_p=1/7$. 

Larger lattice sizes were simulated by using the multicanonical method
as discussed above. We started the detailed study of the transition for
$(\lambda,\mu)=(1.24,-2.3)$,  where we simulated the linear 
lattice sizes $L=8$, $12$, $16$, $24$, $28$, $32$, and $40$. 

In our program no parallelization is implemented. We employed 
trivial parallelization at a moderate level: In the case of 
$L=40$ we performed 5 independent simulations with ordered 
and 5 independent simulations with disordered start configurations
in parallel.
In a series of preliminary runs, as discussed above, we determine 
the weight function $W(X)$ for the multicanonical simulation.

The results given below are based on simulations using our final 
estimate of the weight function. Even when using the multicanonical
simulation, autocorrelation times increase rapidly with increasing 
lattice size. 
In particular in the case of  
larger lattice sizes one has to find a reasonable compromise, when
discarding configurations for equilibration.  We took
$t_{dis} \approx 10 \tau_{ene}$, where $\tau_{ene}$ is the integrated 
autocorrelation time of the energy. We inspected the history 
of our simulations by plotting the expectation values of the energy
or the magnetic susceptibility versus the iteration number of the
Markov chain. We find that with this choice of $t_{dis}$,
a few tunnelings from disorder to order and vice versa are 
discarded. Errors are computed by Jackknife binning with 
$N_{bin}=20$. 
The simulations were performed using a value of $\beta$ slightly
smaller than the preliminary estimate of $\beta_t$ available 
when starting the simulation, giving more weight to the disordered 
phase.  

For $L=40$, we performed for each measurement 30 sweeps with a
local update algorithm and 18 wall cluster updates. With our final
version of $W(X)$, we performed  $5.5 \times 10^7$ measurements after
equilibration.  These simulations took about 120 days on 
a single core of an AMD EPYC$^{TM}$ 7351P CPU. The integrated autocorrelation
time of the energy is about $\tau_{ene} \approx 80000$ in units of measurements.

First we computed the inverse $\beta_t$ of the transition temperature.
To this end, we determined the location $E_{min}$ of the minimum of 
the histogram of the energy density, reweighted to the Boltzmann distribution
for a preliminary estimate of $\beta_t$. Then the estimate of $\beta_t$ is
computed by requiring that the total weight of configurations with 
$E \ge E_{min}$ is $2 N=6$ times as large as that for $E < E_{min}$.
Since the probability density in the neighborhood of $E_{min}$ is very small,  
the estimate of $\beta_t$ is not very sensitive to the exact choice of 
$E_{min}$.  Preliminary analysis shows that replacing the energy 
by, for example, the square of the magnetization leads 
to virtually identical results. Our estimates of $\beta_t$ are summarized 
in table \ref{betatxi2}. 
One expects that $\beta_t$ is converging exponentially fast with increasing
lattice size \cite{BoKo90,BoJa92}. In fact, all estimates obtained for 
$L \ge 12$ are consistent among each other.
As final result we take $\beta_t=0.3294108(5)$, 
obtained for our largest linear lattice size $L=40$. 
For $L=8$,  simulating the
Boltzmann distribution and requiring that $Z_a/Z_p=1/7$ we get
$\beta_t=0.329405(9)$, which is compatible with our final result.

\begin{table}
\caption{\sl \label{betatxi2}
Estimates of the transition temperature $\beta_t$ and the interface 
free energy $F_I$, up to a constant, for a range 
of linear lattice sizes $L$ at $(\lambda,\mu)=(1.24,-2.3)$. 
For a discussion see the text.
}
\begin{center}
\begin{tabular}{rlc}
\hline
 \mc{1}{c}{$L$} &  \mc{1}{c}{$\beta_t$} & $2 F_I + C$ \\
\hline
 8 &  0.329460(25) & 9.12(7) \\    
12 &  0.329409(9)  & 16.30(4) \\   
16 &  0.329405(6)  & 28.01(5) \\  
24 &  0.3294116(20)& 65.07(6) \\  
28 &  0.3294096(19)& 89.01(8) \\   
32 &  0.3294099(9) & 115.76(6) \\  
40 &  0.3294108(5) & 178.90(10) \\ 
\hline
\end{tabular}
\end{center}
\end{table}

\begin{figure}
\begin{center}
\includegraphics[width=14.5cm]{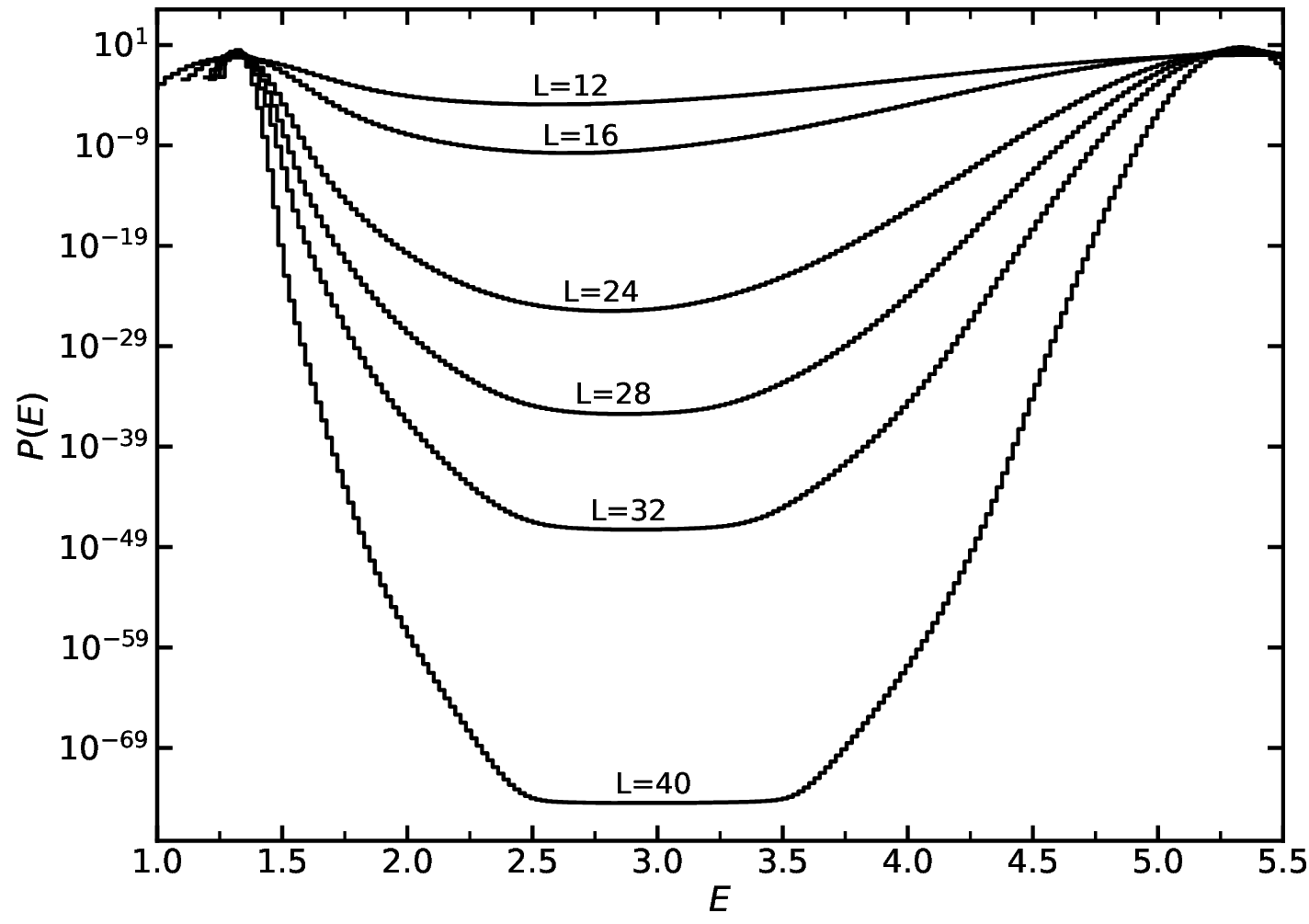}
\caption{\label{histoene}
We give the histograms of the energy density at
$(\lambda,\mu)=(1.24,-2.3)$
for the linear lattice sizes $L=12$, $16$, $24$, $28$, $32$, 
and $40$. The data are reweighted to the Boltzmann distribution 
for $\beta=0.3294108$, which is our 
estimate of the inverse of the transition temperature.
}
\end{center}
\end{figure}

In Fig. \ref{histoene} we plot the histograms for the 
energy density for the Boltzmann distribution at $(\lambda,\mu)=(1.24,-2.3)$,  
$\beta=0.3294108$ and the lattice sizes we have simulated.

Histograms at the first order transition can be understood starting from 
an effective description
of the configuration space. There are regions on the lattice that can 
be assigned to one of the phases. At the transition, all phases have the 
same free energy density. These regions are separated by interfaces, 
which are characterized by their interface tension $\sigma$. The
weight of these sets of configurations is given by the free energy of the
interfaces.

An observable, for example the energy density, takes a certain value 
for each of the phases. There is a characteristic variance
of the observable for each of the phases. The peaks in the histogram
are associated with configurations, where only one phase is present.
To understand the histogram between the peaks we have to consider
configurations, where two phases, the disordered phase and one of the 
ordered phases are present. One has to consider configurations,
which are predominantly associated with one of the phases, and there 
is a droplet of the other phase. Furthermore, on a  $L^3$ lattice with 
periodic boundary conditions, for large $L$, the minimum in the histogram
is related with configurations, where the phases are separated by two flat
interfaces with the area $L^2$.
The reduced free energy of a single interfaces is
\begin{equation}
\label{Gaussian_approx}
 F_I = \sigma L^2 + c \;,
\end{equation}
where the constant $c$ takes into account fluctuations of the interface.
Since $F_I$ does not depend on the distance between the interfaces, 
the histogram becomes flat at the minimum.
Taking into account the translational invariance we get, up to a constant
prefactor,
\begin{equation}
z_{2I}(L) = \exp(2 \ln L - 2 F_I(L))
\end{equation}
as weight for the collection 
of configurations, where two phases are separated by two flat 
interfaces. 
We determine $z_{2I}$ up to a constant prefactor by the value of the 
histogram at its minimum. Our numerical results for 
$2 F_I(L) + C=-\ln(z_{2I}(L))+2 \ln L$
are summarized in table \ref{betatxi2}.

In order to determine the interface tension, we take two lattice sizes 
$L_1$ and $L_2$:
\begin{equation}
\label{sigma_compute}
  \sigma =  \frac{F_I(L_2) - F_I(L_1)}{L_2^2 - L_1^2} \;.
\end{equation}
We arrive at
$\sigma=$
$0.0449(5)$,
$0.0523(3)$, 
$0.0579(1)$,
$0.0566(1)$, and
$0.0548(1)$ for $(L_1,L_2)=(8, 12)$, $(12, 16)$, $(16, 24)$, $(24, 32)$, and
$(32, 40)$. The histograms plotted in Fig. \ref{histoene} show only a clean
plateau value between the peaks for $L=40$ and to a reasonable approximation
for  $L=32$. Therefore we take our estimate obtained for $(L_1,L_2)=(32, 40)$ 
as final result. 

We performed simulations by using the multicanonical method for 
weaker first order phase transitions in a similar
fashion as for $(\lambda,\mu)=(1.24,-2.3)$. The largest lattice 
sizes that we have reached are $L_{max}=64$, $96$, $64$, and $128$ for 
$(\lambda,\mu)=(1.675,-1.95)$, $(2.0, -1.85)$, $(3.0,-2.5)$, and 
$(2.333,-1.764)$, respectively.  Our final results for $\beta_t$ and 
$\sigma$ are given in table \ref{table_xihigh}.
Here our largest $L/\xi_{high}$ are smaller than for 
$(\lambda,\mu)=(1.24,-2.3)$. 
Therefore, systematic errors computing $\sigma$ by using
eq.~(\ref{sigma_compute}) are present. We corrected for that by taking
into account the dependence of the estimate of $\sigma$ seen for 
$(\lambda,\mu)=(1.24,-2.3)$ as a function of $L/\xi_{high}$.

\begin{table}
\caption{\sl \label{table_xihigh}
Final estimates for the first order phase transition. In the first column 
we give the value of the parameters $(\lambda,\mu)$, in the second column
we give the inverse $\beta_t$ of the transition temperature, in the third the 
interface tension for interfaces between the disordered and one of the 
ordered phases. In the fourth column we give the latent heat $Q$ and finally 
in the 
fifth column the correlation length in the high temperature phase. 
For a discussion see the text.
}
\begin{center}
\begin{tabular}{ccccc}
\hline
$(\lambda,\mu)$   & $\beta_t$     & $\sigma$ &  $Q$  &  $ \xi_{high} $ \\
\hline
$(2.333,-1.764)$  & 0.60891779(23)& 0.00046(5)  & 0.04218(15)  & 24.70(2) \\
$(3,-2.5)$        & 0.58194181(50)&    -        & 0.1182(1)  & 12.33(2) \\  
$(2,-1.85)$       & 0.57905753(56)& 0.0019(1)& 0.11681(7) & 12.135(6) \\
$(1.675,-1.95)$   & 0.5306093(64) & 0.0074(4)& 0.3054(5) & 6.003(2) \\  
$(1.24,-2.3)$     & 0.3294108(5)  & 0.055(1) & 1.2012(10) & 1.9833(10) \\ 
\hline
\end{tabular}
\end{center}
\end{table}

\subsubsection{Correlation length}
In order to compute the correlation length in the disordered phase 
at the transition temperature, we simulated the model by using the same program
as above in Sec. \ref{nueff}. The simulations are started with
$\phi_{x,i}= \mbox{rand}-0.5$ for all sites $x$ and components $i$.
The linear lattice size $L$ is chosen such that
the tunneling time to an ordered phase is very large compared with the 
length of the simulation. To this end, we use $L \approx 20 \xi_{high}$, 
self-consistently. The correlation length is determined as
discussed above in Sec. \ref{nueff}. Our final results are summarized 
in table \ref{table_xihigh}. The errors quoted take the uncertainty 
of the inverse transition temperature $\beta_t$ into account.

Computing the correlation length for the ordered phases turns out 
to be considerably more difficult. Here, the connected part of the 
correlation function has to be computed. The improved estimator 
proposed for models with $Z_2$ symmetry \cite{Ha16} can not be applied.
Furthermore, the effective correlation length converges more slowly 
than in the disordered phase. Therefore we computed the 
correlation length for the ordered phases only for  
$(\lambda,\mu)=(1.24,-2.3)$.  We get the rough estimate $\xi_{low}=1.6(1)$. 

\subsubsection{Latent heat}
We define the latent heat as 
\begin{equation}
 Q = \frac{1}{L^3} \left({\langle \cal H} \rangle_{disorder}  - 
\langle {\cal H}\rangle_{order}   \right) 
\end{equation}
taken at the transition temperature $\beta_t$. For the 
disordered phase, the measurements are taken
from a subset of the simulations done for the correlation length discussed
above. For the ordered phases, we performed simulations with the same
lattice sizes  as for the ordered phase. The simulations are started 
with a configuration generated by using eqs.~(\ref{order1},\ref{order2}).
Our results are given in table \ref{table_xihigh}.

\subsubsection{Scaling of the interface tension and the latent heat}
As the first order phase transition becomes weaker, the interface tension
and the latent heat decrease. The combination $\sigma \xi_{high}^2$ should
have a finite limit as the $O(3)$-invariant fixed point is approached. In 
fact we find $\sigma \xi_{high}^2 =  0.216(4)$, $0.267(15)$, $0.280(15)$, 
and $0.28(3)$ for $(\lambda,\mu)=(1.24,-2.3)$, $(1.675,-1.95)$, $(2.0, -1.85)$, 
and $(2.333,-1.764)$, respectively. As our estimate for the scaling limit,
we quote $ \sigma \xi_{high}^2 = 0.28(3)$.

In the case of the latent heat we expect from dimensional analysis that
\begin{equation}
Q \xi_{high}^{d-y_t} = Q \xi_{high}^{\Delta_{\epsilon}} \;,
\end{equation}
where $y_t$ is the thermal RG-exponent of the Heisenberg
universality class, approaches a finite limit as the 
Heisenberg fixed point is approached. 

Taking $y_t=1.4052(2)$ \cite{myIco}, 
we get $Q \xi_{high}^{\Delta_{\epsilon}} = 3.580(4)$, $5.324(9)$, 
$6.256(6)$, $6.494(18)$, and $7.017(27)$
for $(\lambda,\mu)=(1.24,-2.3)$, $(1.675,-1.95)$, $(2.0, -1.85)$, $(3.0, -2.5)$
and $(2.333,-1.764)$, respectively. 
Here the convergence is not as convincing as for the interface tension.
We abstain from quoting a result for the limit $\xi_{high} \rightarrow \infty$.

\section{Summary and conclusion}
\label{summary}
We have studied the three component $\phi^4$ model on the simple cubic lattice 
with a cubic perturbation. Field theory predicts that
the RG-flow rapidly collapses onto a single line in coupling space.
On this line the RG-flow remains slow in an extended range and not just 
in the vicinity of the fixed points. 
See, for example, ref. \cite{AharonyNeu}. Here we corroborate this picture
by using Monte Carlo simulations in connection with a finite size scaling 
analysis. To this end, compared with ref. \cite{myCubic}, we have extended
the study towards stronger breaking of the $O(3)$ invariance.
In a preliminary step, we identify the line of slow flow in the parameters 
$(\lambda,\mu)$
of the reduced Hamiltonian, eq.~(\ref{Hamiltonian}). Here we obtain a more 
accurate characterization than in ref. \cite{myCubic}. Next we study the 
slow RG-flow in the scaling limit. This limit is reached faster for 
$(\lambda,\mu)$ on the line of slow flow than for generic choices. 
In particular at the fixed points, corrections proportional to $L^{-\omega}$
with $\omega \approx 0.8$ should be essentially eliminated. Remaining
corrections are proportional to $L^{-\epsilon_j}$ with 
$\epsilon_j \gtrapprox 2$.  Note that the slow RG-flow 
is not studied in terms of the parameters of the reduced Hamiltonian but
by using the dimensionless quantity $U_C$, eq.~(\ref{UCdef}),
at criticality that quantifies the 
violation of the $O(3)$ symmetry.  Numerically we determine the 
coefficients of the $\beta$-function, 
eqs.~(\ref{infinitesimal_flow},\ref{flowequation},\ref{uAnsatz}), for $U_C$.
The analysis of the $\beta$-function provides us with an accurate estimate 
of the difference 
$Y_4-\omega_2=0.00081(7)$, where $Y_4$ is the RG-exponent of the cubic 
perturbation at the $O(3)$-symmetric fixed point and $\omega_2$ the correction 
exponent at the cubic fixed point.  
Note that for $a_{i \ge 2}=0$, eq.~(\ref{uAnsatz}), $Y_4=\omega_2$.
For example in refs. \cite{AharonyNeu,RoSu23} this approximation of the
$\beta$-function is considered.
Here, analyzing the $\beta$-function
we get $Y_4=0.0141(10)$. In ref. \cite{myCubic}, repeating the FSS analysis
at the $O(3)$-symmetric fixed of ref. \cite{O234} with higher statistics,
we arrived at $Y_4=0.0143(9)$, which is fully consistent. These two approaches
are technically very different, giving us further confidence in the reliability
of the estimates.

In section \ref{nueff}, motivated by ref. \cite{AharonyNeu}, we determine an 
effective exponent $\nu_{eff}$ of the correlation length. First we perform 
a FSS analysis, which provides estimates of $\nu_{eff}$ as a function of $U_C$. 
Then we analyze the behavior of the correlation length in the high temperature
phase for two values of $(\lambda,\mu)$ with $\mu<0$. The results obtained 
by these two different approaches are in reasonable agreement. 

In the final part of the study we focus on the first order phase transition.
For a strong breaking of the $O(3)$ invariance we clearly confirm the 
first order nature of the transition. Histograms of various observables 
show a clear double peak structure. The separation of the two peaks becomes
stronger with increasing lattice size. We obtain accurate estimates of the 
latent heat, the correlation length in the disordered phase at the transition,
and the interface tension of interfaces between the disordered and one of the 
ordered phases. We analyze how these quantities scale with the RG-flow.
This allows us to predict the strength of the transition for parameter
values, where the transition is weak and can therefore not be identified 
as first order directly in the simulations.

The RG-flow numerically studied here is universal. 
Still, applied directly to experiments,
only qualitative and semiquantitative conclusion can be drawn. However,
simulating more realistic models of experimental systems, one could
determine the same dimensionless quantities as discussed here.
In particular computing $U_C$ would connect quantitatively with the 
results obtained here.

\section{Acknowledgement}
This work was supported by the Deutsche Forschungsgemeinschaft (DFG) under
the grant  HA 3150/5-3.

\appendix
\section{Quality of fits and estimating errors}
\label{appendixA}
We perform a standard analysis of our data by using least square fits as
discussed for example in ref. \cite{Young}. Let us assume we have $N$ 
uncorrelated data
points $y_i=y(x_i)$ with a statistical error $\sigma_i$ with a 
Gaussian distribution. These are fitted by using the Ansatz, or model
in the language of statistics, $f(x; p_1,...,p_M)$, where 
$p_j$ are the parameters of the Ansatz. One defines
\begin{equation}
\chi^2 = \sum_i \frac{\left[y_i-f(x_i; p_1,...,p_M)\right]^2}{\sigma_i^2} \;.
\end{equation}
In the case of correlations, given by the covariance matrix $C$, this 
generalizes to 
\begin{equation}
\chi^2 = \sum_{ij} 
\left[y_i-f(x_i,p_1,...,p_M)\right] (C^{-1})_{ij} 
\left[y_j-f(x_j,p_1,...,p_M)\right] \;.
\end{equation}
For given $y_i$ and $\sigma_i$ or $C$, the parameters $p_j$ are obtained by
minimizing $\chi^2$. To this end, we employ the function \verb+curve_fit()+
contained in the SciPy  library \cite{pythonSciPy}. The function
\verb+curve_fit()+ acts as a wrapper to functions contained in the
MINPACK library \cite{MINPACK}. In selected cases, we checked the outcome of the
fit by varying the initial values of the parameters. Furthermore, we performed
fits both by using the Levenberg-Marquardt algorithm and the trust region
reflective algorithm. 
In section \ref{UCflow}, we compute numerically 
$\overline{U}_C^*$, $\omega_2$, and $Y_4-\omega_2$, based on the results for
the parameters $a_0$, $a_1$, ..., $a_n$ of the fit, eq.~(\ref{uAnsatz}).  
In order to obtain the statistical  error of these quantities, we 
compute the partial derivatives of these quantities with respect to the 
statistically 
independent data $y_i$, which are $\overline{U}_C$ for different $(\lambda,\mu)$
and $L$ in the present case. To this end, we repeat the fit $N$ times, where
$N$ is the number of data points. For each of these fits, one of the $y_i$ 
is shifted by a small amount: $\tilde y_i = y_i + \epsilon$ if $i=j$ and
$\tilde y_i = y_i$ else. Then
$
\partial q/\partial y_j \approx (\tilde q_j -q)/\epsilon \;.
$
Here $q$ is the estimate of one of the derived quantities obtained for
the original data, while $\tilde q_j$ is the result obtained for $y_j$ 
shifted.  Then the statistical error $\sigma_q$ is given by
\begin{equation}
 \sigma_q^2 = \sum_{j=1}^{N} \left[\frac{\partial q}{\partial y_j} \right]^2 \; \sigma_j^2 \;.
\end{equation}
To check the correctness of the implementation, we also computed the 
statistical error of the parameters $a_0$, $a_1$, ..., $a_n$ this way, 
giving results consistent with those provided by the function 
\verb+curve_fit()+. Alternatively, one could compute the partial derivatives 
of $q$ with respect to the parameters $a_0$, $a_1$, ..., $a_n$ and then
use the covariance matrix provided by \verb+curve_fit()+ to compute the 
statistical error of $q$.

In order to assess $\chi^2$ we compute
\begin{equation}
p(n_{DOF},\chi^2)= \frac{1}{\Gamma(n_{DOF}/2)} \int_{\chi^2/2}^{\infty} 
t^{n_{DOF}/2-1} \exp(-t) \mbox{d} t \; ,
\end{equation}
where $\Gamma$ is the Euler gamma function and $n_{DOF}=N-M$ the number 
of degrees of freedom. In the Numerical Recipes, Sect. 15 \cite{Numres}
or ref. \cite{Young} this quantity is denoted as goodness of the fit.
The meaning of this quantity is derived for linear Ans\"atze,
see for example Appendix C of ref. \cite{Young}.
For small values of the errors, this should also apply to nonlinear Ans\"atze. 
In the broader context of statistics, the goodness of the fit is the $p$-value 
for the null hypothesis that the Ansatz is correct, without any correction.
Assuming that the null hypothesis is correct, the $p$-value gives the
probability that the data drawn give the value of $\chi^2$ we obtain or 
larger. For small $p$-values one rejects the null hypothesis.
Typically, in the literature, a fit with $p \ge 0.05$ or $0.1$ is
considered to be not excluded. In the text we frequently write sloppily
that the fit is acceptable.

As usual, we perform fits by using a number of different Ans\"atze, with 
a varying degree of approximation compared with the exact form, which
is typically only known in certain limits.
In a first step, these fits are evaluated based on their $\chi^2/$DOF and 
the goodness of the fit obtained from this. Frequently in the literature
a best fit is selected among those that have an acceptable $\chi^2/$DOF and
$p$-value. The parameter values obtained from this best fit are presented 
as the final results, including the statistical error estimates obtained 
by the fit. Implicitly it is assumed that systematic errors, caused by the
imperfection of the Ansatz, are at most of the same size as the statistical
errors. However experience shows that this, depending on the type of 
the approximation, is often not the case and the systematic errors are way
larger. One can easily convince oneself of this fact by generating numbers 
with a known function, put some Gaussian noise on it and then fit these 
numbers by an Ansatz that is an approximation of the function we started
with.

In order to get a better grasp on systematic errors, our final results
are based on a number $n_{fit}$ of fits using different Ans\"atze, with
a varying degree of approximation compared with the exact form. Let us
assume that we get $a_{\alpha}$ with the statistical error
$\sigma_{\alpha}$ for the parameter $a$ of the fits $\alpha=1$, 
$2$,...,$n_{fit}$. Then we quote $(a_{max}+a_{min})/2$ as our final result, 
with the error $(a_{max}-a_{min})/2$, where 
$a_{max}=\mbox{max}_{\alpha} \; [a_{\alpha}+\sigma_{\alpha}]$ and 
$a_{min}=\mbox{min}_{\alpha} \; [a_{\alpha}-\sigma_{\alpha}]$. In the text 
we write that the final result covers the estimates obtained by the 
fits $\alpha=1$, $2$,...,$n_{fit}$, which are taken into account.
Obviously, the final error that is quoted is larger or equal than the 
statistical error of each of the fits. The reliability of the error estimate
depends mainly on the choice of the fits that are taken into account. 
If they all suffer from the major source of systematic error in the 
same way, this error is not reflected in our error estimate. 

Finally let us note that we generated the plots
by using the Matplotlib library \cite{plotting}.

\section{Analytic results for the $\beta$-function}
\label{appendixB}
We discuss the $\beta$-function truncated at third order:
\begin{equation}
\tilde u(x) = a_0 x + a_1 x^2 + a_2 x^3 \; ,
\end{equation}
where for the ease of the notation the coupling is denoted by $x$. 
The indices of $a_i$ are adjusted to the powers in $u(x)=\tilde u(x)/x$. 
The RG fixed points are given by the zeros of $\tilde u(x)$. The first
zero $x=0$ we find trivially. Factoring out this zero, we get
\begin{equation}
\label{uequation} 
u(x) = a_0 + a_1 x + a_2 x^2 \;. 
\end{equation}
Let us first discuss the additional simplification $a_2=0$:
The zero of $u(x)$ is given by $x=-a_0/a_1$. The RG-exponents are 
given by the derivatives of the  $\beta$-function at the zeros.
\begin{equation}
\label{tudiff}
\tilde u'(x) = a_0 + 2 a_1 x + 3 a_2 x^2 \; .
\end{equation}
The RG-exponent at $x=0$ is $Y_4=a_0$, irrespectively from the 
order of the truncation. For $a_i=0$ for $i \ge 2$, we get
for the zero $x=-a_0/a_1$ the RG-exponent
\begin{equation}
\tilde u'(-a_0/a_1) = - \omega_2 = a_0 + 2 a_1 (-a_0/a_1) = - a_0 \;.
\end{equation}
Hence for the truncation $a_i=0$ for $i \ge 2$ we have $Y_4=\omega_2$. 

The solutions of eq.~(\ref{uequation}) for $a_2 \ne 0$ are 
\begin{equation}
 x_{1/2} = \frac{-a_1 \pm \sqrt{a_1^2 - 4 a_0 a_2 }}{2 a_2}   
\end{equation}
In order to get the solution close to  $x=-a_0/a_1$ for small $a_2$, 
we have to select the $+$ from $\pm$. Expanding the square root 
for small $a_2$, we get
\begin{equation}
 x_1= \frac{a_1 \left (- 1 + \sqrt{1 - 4 a_0 a_2/a_1^2} \right)} {2 a_2} 
  =  -  a_0/a_1 - a_2 a_0^2/a_1^3 - ...
\end{equation}
The derivative of $\tilde u(x)$ at this zero is
\begin{eqnarray}
\tilde u'(x_1) &=& 
a_0 + 2 a_1 (-  a_0/a_1 - a_2 a_0^2/a_1^3 - ...) + 3  a_2 (a_0/a_1 + ...)^2 
\nonumber \\
&=& -a_0 - 2 a_2 a_0^2/a_1^2  + 3 a_2 (a_0/a_1)^2 +  ... \nonumber \\
&=& -a_0 + a_2 (a_0/a_1)^2 + ... \;.
\end{eqnarray} 
Hence 
\begin{equation}
Y_0 -  \omega_2 = a_2 (a_0/a_1)^2 + ...  \;.
\end{equation}
For the fits 1, 2, 3, and 4 of section \ref{UCflow}, table \ref{resutable} 
we get $a_2 (a_0/a_1)^2= 0.00063(4)$, $0.00063(2)$, $0.00066(1)$, 
and $0.00066(1)$,
respectively. Hence, numerically $a_2 (a_0/a_1)^2$ contributes more than $3/4$
of the value to $Y_0 -  \omega_2$.

\end{document}